\documentclass[12pt]{article}%
\usepackage[colorlinks,hyperfootnotes=false,pageanchor=false]{hyperref}
\usepackage{amsmath, amsthm}
\usepackage{algorithm}
\usepackage[noend]{algpseudocode}
\usepackage[nameinlink,noabbrev]{cleveref}
\usepackage{tabularx}
\usepackage{tabulary}
\usepackage{placeins}
\usepackage{graphicx}
\usepackage{pgfplots}
\usepackage{tikz}
\usepackage{relsize}
\usepackage{pdflscape}
\usepackage{amssymb}
\usepackage[round]{natbib}
\usepackage{amsfonts}
\usepackage{amsmath}
\usepackage{bm}
\usepackage{latexsym}
\usepackage{epsfig}
\usepackage{rotating}
\usepackage{ctable}
\usepackage{array}
\usepackage{lscape}
\usepackage{float}
\usepackage[left=2.0cm, right=1.8cm, top=1.8cm, bottom=1.7cm]{geometry}%
\providecommand{\U}[1]{\protect\rule{.1in}{.1in}}
\hypersetup{colorlinks=true, linkcolor=black, citecolor=black}
\pgfplotsset{compat=newest}
\pgfplotsset{plot coordinates/math parser=false}
\newlength\figureheight
\newlength\figurewidth

\def\1{1\!{\rm l}}

\newcommand{\ignore}[1]{}

\begin{document}

\title{\textbf{Optimal probabilistic forecasts:\smallskip\ }\\\textbf{When do they work?}\thanks{{\footnotesize This research has been
supported by Australian Research Council (ARC) Discovery Grants DP170100729
and DP200101414. Frazier was also supported by ARC Early Career Researcher
Award DE200101070; and Martin, Loaiza-Maya and Frazier were provided support
by the Australian Centre of Excellence in Mathematics and Statistics.}}}
\author{Gael M. Martin\thanks{{\footnotesize Department of Econometrics and Business
Statistics, Monash University, Australia. Corresponding author:
gael.martin@monash.edu.}}, Rub\'en
Loaiza-Maya\thanks{{\footnotesize Department of Econometrics and Business
Statistics, Monash University, Australia.}}, David T.
Frazier,\thanks{{\footnotesize Department of Econometrics and Business
Statistics, Monash University, Australia.}}
\and Worapree Maneesoonthorn\thanks{{\footnotesize Melbourne Business School,
University of Melbourne, Australia.}} and Andr\'es Ram\'irez
Hassan\thanks{{\footnotesize Department of Economics, Universidad EAFIT,
Colombia.}}\medskip\medskip}
\maketitle

\begin{abstract}
Proper scoring rules are used to assess the out-of-sample accuracy of
probabilistic forecasts, with different scoring rules rewarding distinct
aspects of forecast performance. Herein, we re-investigate the practice of
using proper scoring rules to produce probabilistic forecasts that are
`optimal' according to a given score, and assess when their out-of-sample
accuracy is superior to alternative forecasts, according to that score.
Particular attention is paid to relative predictive performance under
misspecification of the predictive model. Using numerical illustrations, we
document several novel findings within this\ paradigm that\textbf{\ }highlight
the important interplay between the true data generating process, the assumed
predictive model and the scoring rule. Notably, we show that only when a
predictive model is sufficiently compatible with the true process to allow a
particular score criterion to reward what it is designed to reward, will this
approach to forecasting reap benefits. Subject to this compatibility however,
the superiority of the optimal forecast will be greater, the greater is the
degree of misspecification. We explore these issues under a range of different
scenarios, and using both artificially simulated and empirical data.

\bigskip


\bigskip

\emph{MSC2010 Subject Classification}: 60G25, 62M20, 60G35\smallskip


\end{abstract}

%\singlespacing

%\singlespacing

\renewcommand{\baselinestretch}{1}

%\singlespacing

\newpage

\section{Introduction}

\baselineskip18pt

Over the past {two decades}, the use of scoring rules to measure the accuracy
of distributional forecasts has become ubiquitous. In brief, a scoring rule
rewards a probabilistic forecast for assigning a high density ordinate (or
high probability mass) to the observed value, so-called `calibration', subject
to some criterion of `sharpness', or some reward for accuracy in a part of the
predictive support that is critical to the problem at hand. We refer to
\cite{tay2000density}, \cite{gneiting2007probabilistic} and
\cite{Gneiting2007} for early extensive reviews, and \cite{Diks2011} and
\cite{Opschoor2017} for examples of later developments.

In the main, scoring rules have been used to compare the relative predictive
accuracy of probabilistic forecasts produced by different forecasting models
and/or methods. It is fair to say that, on the whole, less attention has been
given to the relationship between the manner in which the forecast is
produced, and the way in which its accuracy is assessed. Exceptions to this
comment include \cite{gneiting2005calibrated}, \cite{Gneiting2007},
\cite{elliott2008economic}, {\cite{loaiza2019focused}} {and }%
\cite{patton2019comparing}, and\ related work on the scoring of point,
quantile or expectile forecasts
in\ \cite{gneiting2011quantiles,gneiting2011making}, {\cite{holzmann2014}},
\cite{ehm2016quantiles}, \cite{fissler2016higher}, \cite{kruger:2020} {and}
\cite{ziegel2020robust}. In this work, focus is given to producing forecasts
that are, in some sense, \textit{optimal }for the particular empirical problem
and - as part of that - deliberately matched to the score used to evaluate
out-of-sample performance; the\textbf{\ }idea here being that the forecast so
chosen will, by construction, perform best out-of-sample according that
scoring rule. The literature on optimal forecast \textit{combinations} is
similar in spirit, with the combination weights chosen with a view to
optimizing a particular forecast-accuracy criterion. (See
\citealt{aastveit2018evolution}, for a recent review.)

Our work continues in this vein, but with three very specific, and
inter-related questions addressed regarding the production of an `optimal'
probabilistic forecast via the optimization of a criterion function defined by
a\textbf{\ }given scoring rule. First, what is the impact of model
misspecification on the performance of an optimal forecast? Second, when can
we be assured that an optimal forecast \textit{will} yield the best
out-of-sample performance, as measured by the relevant score? Third, when can
we have no\textit{\ }such assurance?

We phrase answers to these questions in terms of the concept of `coherence': a
forecast that is optimal with respect to a particular score is said to be
`coherent' if it cannot be beaten out-of-sample (as assessed by that score) by
a forecast that is optimal according to a different score.\textbf{\ }An
optimal forecast is `strictly coherent' if it is strictly preferable to all
alternatives, when evaluated according to its own score. The word `coherent'
is used here to reflect the fact that the method chosen to produce a forecast
performs out-of-sample in a manner that fits with, or is \textit{coherent
}with\ that choice: i.e. no other choice (within the context of forecasts
produced via proper scoring rules) is strictly preferable.\textbf{\ }

Nested within this concept of coherence is the known result
(\citealp{Gneiting2007}; \citealp{patton2019comparing}) that correct
specification of the model, and under equivalent conditioning sets,\textbf{\ }%
leads to optimal forecasts that have theoretically equivalent out-of-sample
performance according to any proper score, with numerical differences
reflecting sampling variation only. That is, all\ such methods are coherent in
the sense that,\textbf{\ }in the limit, no one forecast is out-performed by
another. However, the concept of coherence really has most import in the
empirically relevant\textbf{\ }case where a predictive model is misspecified.
In this setting, one cannot presume that estimating the parameters of a
predictive model by optimizing any proper criterion will reveal the true
predictive model. Instead, one is forced to confront the fact that no such
`true model' will be revealed, and that the criterion should be defined by a
score that rewards the type of predictive accuracy that matters for the
problem at hand. It is in this misspecified setting that we would hope to see
strict\textbf{\ }coherence on display; providing justification as it would for
simply producing a forecast via the scoring rule that is pertinent to the
problem at hand, and leaving matters at that.\footnote{We note that throughout
the paper we only consider examples in which there is a single, common
conditioning set. That is, in contrast to \cite{holzmann2014} and\textbf{\ }%
\cite{patton2019comparing} for example, we do not explore the impact on
{relative predictive performance} of different conditioning sets.}

The concept of `coherence' is distinct from the concept of `consistency' that
is used in some of the literature cited above (e.g.
\citealp{gneiting2011making}, \citealp{holzmann2014},
\citealp{ehm2016quantiles}, and \citealp{patton2019comparing}). As pointed out
by \citeauthor{patton2019comparing}, in the probabilistic forecasting setting
a `consistent' scoring function is analogous to a `proper' scoring rule, which
is `consistent' for the true forecast distribution in the sense of being
maximized (for positively-oriented scores) at that distribution. We restrict
our attention \textit{only} to proper (or `consistent') scores. Within that
set of scores, we then document when optimizing according to any \textit{one}
proper score produces out-of-sample performance - according to that\textit{\ }%
score - that is superior to that of predictions deduced by optimizing
alternative scores, and when it does not; i.e. when \textit{strict coherence}
between in-sample estimation and out-of-sample performance is in evidence and
when it is not. What we illustrate is that the extent to which coherent
forecasts arise in practice actually depends on the form, and degree of
misspecification. First, if the interplay between the predictive model and the
true data generating process is such that a particular score cannot reward the
type of predictive accuracy it is\textbf{\ }designed to reward,\textbf{\ }then
optimizing that model according to that score will not necessarily lead to a
\textit{strictly} coherent forecast. Second, if a misspecified model is
sufficiently `compatible' with the process generating the data, in so much as
it allows a particular score criterion to reward what it was designed to,
strict coherence will indeed result; with the {superiority} of the optimal
forecast being more marked, the greater the degree of misspecification,
subject to this basic compatibility.

We demonstrate all such behaviours in the context of both probabilistic
forecasts based on a single parametric model, and forecasts produced by a
\textit{linear} \textit{combination}\textbf{\ }of predictive distributions. In
the first case optimization is performed with respect to the parameters of the
assumed model; in the second case optimization is with respect to both the
weights of the linear combination and the parameters of the constituent
predictives. To reflect our focus on model misspecification, at no point do we
assume that the true model is spanned by the linear pool; that is, we adopt
the so-called\textbf{\ }$\mathcal{M}$-open view of the world
(\citealt{bernardo1994bayesian}). Our results contribute to the active
literature on frequentist estimation of linear predictive combinations via
predictive criteria (e.g. {\citealp{Hall2007}, \citealp{ranjan2010},
\citealp{Clements2011}, \citealp{Geweke2011}, \citealp{gneiting:2013},
\citealp{Kap2015}, \citealp{Opschoor2017}, \citealp{ganics2018optimal} and
\citealp{pauwels2020higher}). In particular, our results provide a possible
explanation behind the often mixed out-of-sample performance of optimal
weighting schemes. }

{After introducing the concept of coherent predictions in Section \ref{coh},
in Section \ref{numerics} w}e conduct a set of simulation exercises, with a
range of numerical and graphical results used to illustrate coherence
(including strict incoherence) under various design scenarios. Attention is
given to accurate prediction of extreme values of financial returns, by using
- as the optimization criterion - a scoring rule that rewards accurate
prediction in the tails of the predictive distribution. We provide a simple
example that illustrates how easy it is to stumble upon a model that lacks
sufficient compatibility with the true data generating process to prevent a
nominally `optimal' forecast from out-performing others in the manner
expected. The illustration using linear pools highlights the fact that optimal
pools \textit{can} reap benefits relative to alternative\textbf{\ }approaches.
However, the very use of a combination of predictives to provide a more
flexible and, hence, less misspecified representation of the true model can in
some cases mitigate against the benefits of optimization. Section \ref{emp}
{documents the results of an }empirical exercise that focuses on accurate
prediction of {returns on} the S\&P500 index and the MSCI Emerging Market
(MSCIEM) index. Once again we demonstrate that there are benefits in seeking a
predictor that is optimal according to a particular scoring rule, with
slightly\textbf{ }more marked gains in evidence in the case of the single
predictive models than in the case of the linear pool. The paper concludes in
Section \ref{disc}.

\section{Optimal Prediction and the Notion of Coherence\label{coh}}

\subsection{Optimal Prediction}

Let $(\Omega,\mathcal{F},G_{0})$ be a probability space, and let
$Y_{1}^{\infty}:=\{Y_{1},\dots,Y_{n},\dots\}$ be a sequence of random
variables whose infinite-dimensional distribution is $G_{0}$. In general,
$G_{0}$ is unknown, and so a hypothetical class of probability distributions
is postulated for $G_{0}$. Let $\mathbb{P}$ be a convex class of probability
distributions operating on $(\Omega,\mathcal{F})$ that represents our best
approximation of $G_{0}$.

Assume our goal is to analyze the ability of the distribution $P\in\mathbb{P}$
{to generate} accurate probabilistic forecasts. The most common concept {used
}to capture accuracy of such forecasts is a scoring rule. A scoring rule is a
function $S:\{\mathbb{P}\cup\{G_{0}\}\}\times\Omega\mapsto\mathbb{R}$ whereby
if the forecaster quotes the distribution $P$ and the value $y$ eventuates,
then the reward (or `score') is $S(P,y).$ {As described earlier, in general
terms a s}coring rule rewards a forecast for assigning a high density ordinate
(or high probability mass) {to }$y$, often subject to some shape, {or
sharpness} criterion, with higher scores denoting qualitatively better
predictions {than lower scores}, assuming all scores are positively-oriented.

The result of a single score evaluation is, {however,} of little use by itself
as a measure of predictive accuracy. To obtain a {meaningful gauge} of
predictive accuracy, as measured by $S(\cdot,\cdot)$, we require some notion
of regularity against which different {predictions can be assessed}. {By far
the }most common measure of regularity used in the literature is via the
notion of {the \textit{expected} score}: following \cite{Gneiting2007}, the
expected score under the true measure $G_{0}$\ of the probability forecast
$P$, is{\ given by }$\mathbb{S}(P,G_{0})=\int_{y\in\Omega}S(P,y)dG_{0}(y).$ A
scoring rule $S(\cdot,\cdot)$ is `proper' relative to\textbf{ }$\mathbb{P}$ if
for all $P,G_{0}\in\mathbb{P}$, $\mathbb{S}(G_{0},G_{0})\geq\mathbb{S}%
(P,G_{0}),$ and is\textbf{\ }\textit{strictly }proper\ if $\mathbb{S}%
(G_{0},G_{0})=\mathbb{S}(P,G_{0})\Leftrightarrow$ $P=G_{0}$. That is, if the
forecaster's best judgement is $G_{0}$, then a proper scoring rule rewards the
forecaster for quoting $P=G_{0}$.

The {concept} of a proper scoring rule is useful {from a practical perspective
}since it guarantees that, if we knew that the true DGP was $G_{0}$,
then\textbf{\ }according to the rule $S(\cdot,\cdot)$ the best forecast we
could hope to obtain would, on average, result by choosing $G_{0} $. Note that
this notion of `average'\ is embedded into the very definition of a proper
scoring rule since it, itself, relies on the notion of an\ expectation. It is
clear that, {in practice,} the expected score $\mathbb{S}(\cdot,G_{0})$ is
unknown and {cannot }be calculated. However, if one believes that the true DGP
is {an element of} $\mathbb{P}$, {a sensible approach to adopt is to} {form
}an empirical version of $\mathbb{S}(\cdot,G_{0})$ and {search} $\mathbb{P}$
to find the `best'\ predictive over this\textbf{\ }class \citep{Gneiting2007}.

More formally, for $\tau$ such that $T\geq\tau\geq1$, let $\{y_{t}%
\}_{t=2}^{T-\tau}$ denote a series of size\textbf{\ }$T-(\tau+1)$, over which
we wish to search for the most accurate predictive, where $T$ is the total
number of observations on\textbf{\ }$y_{t}$, and where $\tau$ denotes the size
of a {hold-out sample}. Assume that the class of models under analysis is
indexed by a vector of unknown parameters $\boldsymbol{\theta}\in\Theta
\subset\mathbb{R}^{d_{\boldsymbol{\theta}}}$, i.e., $\mathbb{P}\equiv
\mathbb{P}(\Theta)$,\textbf{\ }where $\mathbb{P}(\Theta):=\{\boldsymbol{\theta
}\in\Theta:P_{\boldsymbol{\theta}}\}.$ For $\mathcal{F}_{t-1}$ denoting the
time $t-1$ information set, and for each $\boldsymbol{\theta}\in\Theta$, we
associate to the model $P_{\boldsymbol{\theta}}$ the predictive measure
$P_{\boldsymbol{\theta}}^{t-1}:=P(\cdot|\mathcal{F}_{t-1},\boldsymbol{\theta
})$ and, where applicable, the associated predictive density $p(\cdot
|\mathcal{F}_{t-1},\boldsymbol{\theta})$.{\ We can define an estimator
of\ }$\boldsymbol{\theta}$ {as}
\begin{equation}
\widehat{\boldsymbol{\theta}}:=\arg\max_{\boldsymbol{\theta}\in\Theta
}\overline{S}(\boldsymbol{\theta}), \label{opt_theta}%
\end{equation}
where%
\begin{equation}
\overline{S}(\boldsymbol{\theta}):=\frac{1}{T-(\tau+1)}\sum_{t=2}^{T-\tau
}S\left(  P_{\boldsymbol{\theta}}^{t-1},y_{t}\right)  , \label{opt}%
\end{equation}
and where the notation $\overline{S}(.)$ clarifies that the criterion function
is a sample average, with components defined by the particular choice of
score, $S(\cdot,\cdot).$ The estimator $\widehat{\boldsymbol{\theta}}$ is
referred to as the \textit{optimal score estimator} for the scoring rule
$S(\cdot,\cdot)$, and $P_{\widehat{\boldsymbol{\theta}}}^{t-1}$ as the
\textit{optimal predictive}.

\subsection{Coherent Predictions\label{coherent}}

The predictive $P_{\widehat{\boldsymbol{\theta}}}^{t-1}$ is `optimal' in the
following specific sense:{\ if our goal is to achieve good forecasting
performance according {to }the given scoring rule $S$, all we have to do is
optimize the parameters of the {predictive} according to this rule. Implicitly
then, if we have two proper scoring rules $S_{1}$ and $S_{2}$, by which we
}produce{\ two different {optimal} predictives }%
$P_{\widehat{\boldsymbol{\theta}}_{1}}^{t-1}${\ and }%
$P_{\widehat{\boldsymbol{\theta}}_{2}}^{t-1}$,{\ where
$\widehat{\boldsymbol{\theta}}_{1}$ and $\widehat{\boldsymbol{\theta}}_{2}$
denote the optimizers according to $S_{1},S_{2}$,}\textbf{\ }it should be the
case that, for large enough $\tau$,%
\begin{equation}
\frac{1}{\tau}\sum_{t=T-\tau+1}^{T}S_{1}\left(  P_{\widehat{\boldsymbol{\theta
}}_{1}}^{t-1},y_{t}\right)  \geq\frac{1}{\tau}\sum_{t=T-\tau+1}^{T}%
S_{1}\left(  P_{\widehat{\boldsymbol{\theta}}_{2}}^{t-1},y_{t}\right)
\label{1}%
\end{equation}
and
\begin{equation}
\frac{1}{\tau}\sum_{t=T-\tau+1}^{T}S_{2}\left(  P_{\widehat{\boldsymbol{\theta
}}_{1}}^{t-1},y_{t}\right)  \leq\frac{1}{\tau}\sum_{t=T-\tau+1}^{T}%
S_{2}\left(  P_{\widehat{\boldsymbol{\theta}}_{2}}^{t-1},y_{t}\right)  .
\label{2}%
\end{equation}
{That is, \textit{coherent} results are expected: }the {predictive that is
optimal with respect to }$S_{1}${\ cannot be beaten out-of-sample (as assessed
by that score) by a predictive that is optimal according to }$S_{2}$, {and
vice-versa. }As mentioned earlier, this definition of coherence subsumes the
case where\textbf{\ }$G_{0}\in\mathbb{P}$ and, $\widehat{\boldsymbol{\theta}%
}_{1}$ and $\widehat{\boldsymbol{\theta}}_{2}$ are both consistent for the
true (vector) value of\textbf{\ }$\boldsymbol{\theta}$, $\boldsymbol{\theta
}_{0}$.\textbf{\ }Hence, in this special case, for\textbf{ $\tau
\rightarrow\infty$,} the expressions in (\ref{1}) and (\ref{2}) would collapse
to equalities.

What is of relevance empirically though, as already highlighted, is the case
where $G_{0}\notin\mathbb{P}.$ Whether coherence holds in this setting depends
on four things: the unknown true model, $G_{0}$, the assumed (but
misspecified) model, $P_{\boldsymbol{\theta}}$, and the two rules,
$S_{1},S_{2}$, in which we are optimizing to obtain predictive distributions.
As we will illustrate with particular examples, it is this collection,
$\{G_{0},P_{\boldsymbol{\theta}},S_{1},S_{2}\}$, that\textbf{\ }determines
whether or not the above notion of coherence holds.

We begin this illustration with a series of simulation experiments in Section
\ref{numerics}. We first specify a single predictive model
$P_{\boldsymbol{\theta}}$ to be, in order: correctly specified; misspecified,
but suitably `compatible' with $G_{0}$ to allow \textit{strict}
coherence\textbf{\ }to prevail; and misspecified in a way in which strict
coherence does not hold; where by \textit{strict} coherence we mean that
\textit{strict inequalities} hold in expressions like (\ref{1}) and (\ref{2}).
The numerical results are presented in a variety of different ways, in order
to shed light on this phenomenon of coherence and help practitioners gain an
appreciation of what they should be alert to. As part of this exercise, we
make the link between our broadly descriptive analysis and the formal test of
equal predictive ability of \cite{giacomini2006tests}, in a manner to be
described. Six different proper scoring rules are entertained - both in the
production of the optimal predictions and in the out-of-sample evaluation. We
then shift the focus to a linear predictive pool that does not span the true
model and is, as a consequence, misspecified; documenting the nature of
coherence in this context. In Section \ref{emp} the illustration - based on
both single models and linear pools - proceeds with empirical returns data.

\section{Simulation Experiments\label{numerics}}

\subsection{Single Model Case: Simulation Design\label{single_design}}

In this first set of simulation experiments the aim is to produce an optimal
predictive distribution for a variable that possesses the stylized features of
a financial return. With this in mind, we assume a predictive associated with
an autoregressive conditional heteroscedastic model of order 1 (ARCH(1)) for
the logarithmic return, $y_{t}$,%
\begin{equation}
y_{t}=\phi_{1}+\sigma_{t}\varepsilon_{t};\text{ }\sigma_{t}^{2}=\phi_{2}%
+\phi_{3}\left(  y_{t-1}-\phi_{1}\right)  ^{2};\text{ }{\varepsilon_{t}\sim
i.i.d.N\left(  0,1\right)  .} \label{arch_1}%
\end{equation}
{Panels A, B and C of }Table \ref{tab:desing} then describe both the true data
generating process (DGP) and the precise specification of the model in
(\ref{arch_1}) for the three scenarios:{\ correct specification in (i), and
two different types of misspecification in (ii) and (iii)}. As is clear: in
scenario (i), the assumed model matches the Gaussian ARCH(1) model that has
generated the data; in scenario (ii) - in which a generalized ARCH (GARCH)
model with Student $t$ innovations generates $y_{t}$ - it does not; whilst in
scenario (iii) there is not only misspecification of the assumed model, but
the marginal mean in that model is held fixed at zero. Thus, in the third
case, the predictive of the assumed model is unable to shift location and,
hence, to `move' to accommodate extreme observations, in either tail. The
consequences of this, in terms of relative predictive accuracy, are
highlighted below.

\begin{table}[h]
\caption{Simulation designs for the single model case. Here, $t_{\nu}$ denotes
the Student $t$ distribution with $\nu$ degrees of freedom.\medskip}%
\label{tab:desing}%
\centering%
\begin{tabular}
[c]{lccc}\hline\hline
&  &  & \\
& Panel A & Panel B & Panel C\\\cline{0-0}\cline{2-4}
&  &  & \\
& \underline{Scenario (i)} & \underline{Scenario (ii)} & \underline{Scenario
(iii)}\\
&  &  & \\
& $y_{t}=\sigma_{t}\varepsilon_{t}$ & $y_{t}=\sqrt{\frac{\nu-2}{\nu}}%
\sigma_{t}\varepsilon_{t}$ & $y_{t}=\sqrt{\frac{\nu-2}{\nu}}\sigma
_{t}\varepsilon_{t}$\\
True DGP & $\sigma_{t}^{2}=1+0.2y_{t-1}^{2}$ & $\sigma_{t}^{2}=1+0.2y_{t-1}%
^{2}+0.7\sigma_{t-1}^{2}$ & $\sigma_{t}^{2}=1+0.2y_{t-1}^{2}+0.7\sigma
_{t-1}^{2}$\\
& $\varepsilon_{t}\sim i.i.d.N(0,1)$ & $\varepsilon_{t}\sim i.i.d.t_{\nu}$ &
$\varepsilon_{t}\sim i.i.d.t_{\nu}$\\
&  &  & \\
& $y_{t}=\theta_{1}+\sigma_{t}\varepsilon_{t}$ & $y_{t}=\theta_{1}+\sigma
_{t}\varepsilon_{t}$ & $y_{t}=\sigma_{t}\varepsilon_{t}$\\
Assumed model & $\sigma_{t}^{2}=\theta_{2}+\theta_{3}(y_{t-1}-\theta_{1})^{2}$
& $\sigma_{t}^{2}=\theta_{2}+\theta_{3}(y_{t-1}-\theta_{1})^{2}$ & $\sigma
_{t}^{2}=\theta_{2}+\theta_{3}y_{t-1}^{2}$\\
& $\varepsilon_{t}\sim i.i.d.N(0,1)$ & $\varepsilon_{t}\sim i.i.d.N(0,1)$ &
$\varepsilon_{t}\sim i.i.d.N(0,1)$\\
&  &  & \\\hline\hline
\end{tabular}
\end{table}The predictive $P_{\boldsymbol{\theta}}^{t-1}$, with density
$p\left(  y_{t}|\mathcal{F}_{t-1},\boldsymbol{\theta}\right)  $, is associated
with the assumed Gaussian ARCH(1) model, where $\boldsymbol{\theta}%
=(\theta_{1},\theta_{2},\theta_{3})^{\prime}$. We estimate $\boldsymbol{\theta
}$ as in (\ref{opt_theta}) using the following three types of scoring
rules:{\textbf{\ }}for $I(y\in A)${\textbf{\ }}the indicator on the event
$y\in A$,{\
\begin{align}
S_{\text{LS}}(P_{\boldsymbol{\theta}}^{t-1},y_{t})  &  =\ln p\left(
y_{t}|\mathcal{F}_{t-1},\boldsymbol{\theta}\right)  ,\label{ls_prelim}\\
S_{\text{CRPS}}(P_{\boldsymbol{\theta}}^{t-1},y_{t})  &  =-\int_{-\infty
}^{\infty}\left[  P\left(  y|\mathcal{F}_{t-1},\boldsymbol{\theta}\right)
-I(y\geq y_{t})\right]  ^{2}dy,\label{crps}\\
S_{\text{CLS}}(P_{\boldsymbol{\theta}}^{t-1},y_{t})  &  =\ln p\left(
y_{t}|\mathcal{F}_{t-1},\boldsymbol{\theta}\right)  I\left(  y_{t}\in
A\right)  +\left[  \ln\int_{A^{c}}p\left(  y|\mathcal{F}_{t-1}%
,\boldsymbol{\theta}\right)  dy\right]  I\left(  y_{t}\in A^{c}\right)  ,
\label{csr_prelim}%
\end{align}
}where $P\left(  .|\mathcal{F}_{t-1},\boldsymbol{\theta}\right)  $ in
(\ref{crps}) denotes the predictive cumulative distribution function
associated with $p\left(  .|\mathcal{F}_{t-1},\boldsymbol{\theta}\right)  $.
{Use of the} log-score (LS) in (\ref{ls_prelim})\textbf{\ }{yields the average
log-likelihood function as the criterion in (\ref{opt}) and, under correct
specification and appropriate regularity, the asymptotically efficient
estimator of }$\boldsymbol{\theta}_{0}.$ The score in (\ref{csr_prelim}) is
the\textbf{ }censored likelihood score (CLS)\ of \cite{Diks2011}.{ This score
rewards predictive accuracy over any region of interest $A$ ($A^{c}%
${\ denoting the complement of this region). We report results for }$A$
{defining} the lower {and upper tails of the predictive distribution, as
determined in turn by the 10\%, 20\%, 80\% and 90\% {percentiles}}{\ of the
empirical distribution of }$y_{t}${. The results based on the use of
(\ref{csr_prelim}) in (\ref{opt}) }are labelled hereafter {as CLS }$10\%$,
{CLS }$20\%,$ {CLS }$80\%$ and {CLS }$90\%.$ }{The continuously ranked
probability score (}CRPS) (see \citealp{Gneiting2007}) is sensitive to
distance,\ and rewards the assignment of high predictive mass \textit{near to}
the realized value of $y_{t}$, rather just \textit{at }that value, as in the
case of the log-score. It can be evaluated in closed form for the
(conditionally) Gaussian predictive model assumed under all three scenarios
described in Table \ref{tab:desing}.\ Similarly, in the case of the CLS in
(\ref{csr_prelim}), all components, including the integral $\int_{A^{c}%
}p\left(  y|\mathcal{F}_{t-1},\boldsymbol{\theta}\right)  dy$, have
closed-form representations for the Gaussian predictive model. Note that all
scores are {positively-oriented}; hence, higher values indicate greater
predictive accuracy.\medskip

For each of the Monte Carlo designs, we conduct the following steps:

\begin{enumerate}
\item Generate $T$ observations of $y_{t}$ from the true DGP;

\item Use observations $t=1,...,1,000$ to compute $\widehat{\boldsymbol{\theta
}}^{[i]}$ as in (\ref{opt_theta}), for $S_{i}$, $i\in$ \{LS, CRPS, CLS 10\%,
CLS 20\%, CLS 80\%, CLS 90\%\};

\item Construct the one-step-ahead predictive $P_{\widehat{\boldsymbol{\theta
}}^{[i]}}^{t-1}$, and compute the score, \newline$S_{j}\left(
P_{\widehat{\boldsymbol{\theta}}^{[i]}}^{t-1},y_{t}\right)  $, based on the
`observed' value, $y_{t}$, using $S_{j}$, $j\in$ \{LS, CRPS, CLS 10\%, CLS
20\%, CLS 80\%, CLS 90\%\};

\item Expand the estimation sample by one observation and repeat Steps 2 and
3, retaining\textbf{\ }notation\textbf{\ }$\widehat{\boldsymbol{\theta}}%
^{[i]}$\textbf{\ }for the $S_{i}$-{based }estimator of\textbf{\ }%
$\boldsymbol{\theta}$ {constructed from} each expanding sample.\textbf{\ }Do
this $\tau=T-1,000$ times, and compute:%
\begin{equation}
\overline{S}_{j}(\widehat{\boldsymbol{\theta}}^{[i]})=\frac{1}{\tau}%
\sum_{t=T-\tau+1}^{T}S_{j}\left(  P_{\widehat{\boldsymbol{\theta}}^{[i]}%
}^{t-1},y_{t}\right)  \label{score_ij}%
\end{equation}
for each\textbf{\ }$(i,j)$\textbf{\ }combination.
\end{enumerate}

The results are tabulated and discussed in Section \ref{scores}.

\subsection{Single Model Case: Simulation Results\label{scores}}

In Table \ref{tab:simgarchTrueSpec}, results are recorded for both
$\tau=5,000$ and $\tau=10,000$, and under correct specification of the
predictive model. The large value of\textbf{\ }$\tau=5,000$ is adopted in
order to minimize the effect of sampling error on the results. The even larger
value of\textbf{\ }$\tau=10,000$\textbf{\ }is then adopted as a check
that\textbf{\ }$\tau=5,000$ is \textit{sufficiently} large to be used in all
subsequent experiments. All numbers on the main diagonal correspond to
$\overline{S}_{j}(\widehat{\boldsymbol{\theta}}^{[i]})$ in (\ref{score_ij})
with $i=j.$ Numbers on the off-diagonal correspond to $\overline{S}%
_{j}(\widehat{\boldsymbol{\theta}}^{[i]})$ with $i\neq j.$ Rows in the table
correspond to the $ith$ optimizing criterion (with
$\widehat{\boldsymbol{\theta}}^{[i]}$ the corresponding `optimizer'), and
columns to the results based on the\textbf{\ }$jth$ out-of-sample score,
$S_{j}$. Using the definition of coherence in Section \ref{coherent}, and
given the correct specification, we would expect any given diagonal element to
be equivalent\textit{ }to all values in the column in which it appears, at
least up to sampling error. As is clear, in Panel A the results based on
$\tau=5,000\,$\ essentially bear out this expectation; in Panel B, in which
$\tau=10,000$, all but three results display the requisite equivalence, to two
decimal places.

In Tables \ref{tab:simgarchnu3} and \ref{tab:simgarchnu3_fm} we record results
for the misspecified designs. Given the close correspondence between the
$\tau=5,000$ and $\tau=10,000$ results in the correct specification case, we
now record results based on $\tau=5,000$\ only. In Table \ref{tab:simgarchnu3}%
, the degrees of freedom in the Student $t$ innovation of the true DGP moves
from being very low ($\nu=3$ in Panel A) to high ($\nu=30$ in Panel C),
thereby producing a spectrum of misspecification - {at least in terms of the
distributional form of the innovations -} from very severe to{\ less severe};
and the results on relative out-of-sample accuracy change accordingly. In
Panel A, a \textit{strict }form of coherence is in evidence:\textbf{\ }each
diagonal value exceeds all other values in its column (and is highlighted in
bold accordingly). As $\nu$ increases, the diagonal values remain bold,
although there is less difference between the numbers in any particular
column. Hence, in this case, the advice to a practitioner would certainly be
to optimize the score criterion that is relevant. In particular, given the
importance of accurate estimation of extreme returns, the edict would indeed
be: produce a predictive based on an estimate of $\boldsymbol{\theta}$ that is
optimal in terms of the relevant CLS-based criterion. Given{\ the chosen
predictive model, no other estimate of this model} will produce better
predictive accuracy in the relevant tail, and {this specific estimate} may
well yield quite markedly superior results {to} any other choice, depending on
the fatness of the tails in the true DGP.

In Table \ref{tab:simgarchnu3_fm} however, the results tell a {very}%
{\ }different story. In particular, in Panel A - despite $\nu$\ being very low
- with one exception (the optimizer based on CRPS), the predictive based on
any given optimizer is \textit{never }superior out-of-sample according to that
same score criterion; i.e. the main diagonal is not uniformly diagonal. A
similar comment applies to the results in Panels B and C. In other words, the
assumed predictive model - in which the marginal mean is held fixed - is not
flexible enough to allow any particular scoring rule to produce a point
estimator that delivers good out-of-sample performance in that rule. For
example, the value of $\boldsymbol{\theta}$ {that optimizes the criterion in
(\ref{opt}) based on CLS}{{\ }$10\%$ {does not} }{correspond to an estimated
predictive that gives a high score to extremely low values of }$y_{t}$, {as
the predictive model cannot shift location, and thereby assign high density
ordinates to these values. The assumed model is, in this sense, incompatible
with the true DGP, which will sometimes produce very low values of }$y_{t}.$

\begin{table}[ptb]
\caption{{\protect\footnotesize Average out-of-sample scores under a correctly
specified Gaussian ARCH (1) model (Scenario (i) in Table 1). Panel A (B)
reports the average scores based on }$\tau=5,000${\protect\footnotesize \ (}%
$\tau=10,000$) {\protect\footnotesize out-of-sample values. The rows in each
panel refer to the optimizer used. The columns refer to the out-of-sample
measure used to compute the average scores. The figures in bold are the
largest average scores according to a given out-of-sample measure.\medskip}}%
\label{tab:simgarchTrueSpec}%
\centering%
\begin{tabular}
[c]{lrrrrrr}\hline\hline
&  &  &  &  &  & \\
& \multicolumn{6}{c}{\textbf{Panel A: 5,000 out-of-sample evaluations}}\\
&  &  &  &  &  & \\
& \multicolumn{6}{c}{\textbf{Out-of-sample score}}\\\cline{2-7}
&  &  &  &  &  & \\
& \multicolumn{1}{l}{LS} & \multicolumn{1}{l}{CRPS} & \multicolumn{1}{l}{CLS
10\%} & \multicolumn{1}{l}{CLS 20\%} & \multicolumn{1}{l}{CLS 80\%} &
\multicolumn{1}{l}{CLS 90\%}\\
\textbf{Optimizer} &  &  &  &  &  & \\\cline{1-1}
&  &  &  &  &  & \\
LS & \textbf{-1.510} & \textbf{-0.624} & \textbf{-0.376} & \textbf{-0.602} &
\textbf{-0.593} & \textbf{-0.363}\\
CRPS & \textbf{-1.510} & \textbf{-0.624} & \textbf{-0.376} & \textbf{-0.602} &
\textbf{-0.593} & \textbf{-0.363}\\
CLS 10\% & -1.512 & -0.625 & -0.377 & \textbf{-0.602} & -0.595 & -0.365\\
CLS 20\% & -1.514 & -0.625 & -0.377 & \textbf{-0.602} & -0.597 & -0.366\\
CLS 80\% & -1.518 & -0.626 & -0.381 & -0.609 & -0.594 & \textbf{-0.363}\\
CLS 90\% & -1.516 & -0.626 & -0.380 & -0.608 & -0.594 & \textbf{-0.363}%
\\\hline\hline
&  &  &  &  &  & \\
& \multicolumn{6}{c}{\textbf{Panel B: 10,000 out-of-sample evaluations}}\\
&  &  &  &  &  & \\
& \multicolumn{6}{c}{\textbf{Out-of-sample score}}\\\cline{2-7}
&  &  &  &  &  & \\
& \multicolumn{1}{l}{LS} & \multicolumn{1}{l}{CRPS} & \multicolumn{1}{l}{CLS
10\%} & \multicolumn{1}{l}{CLS 20\%} & \multicolumn{1}{l}{CLS 80\%} &
\multicolumn{1}{l}{CLS 90\%}\\
\textbf{Optimizer} &  &  &  &  &  & \\\cline{1-1}
&  &  &  &  &  & \\
LS & \textbf{-1.510} & \textbf{-0.623} & \textbf{-0.367} & \textbf{-0.598} &
\textbf{-0.597} & \textbf{-0.364}\\
CRPS & \textbf{-1.510} & \textbf{-0.623} & \textbf{-0.367} & \textbf{-0.598} &
\textbf{-0.597} & \textbf{-0.364}\\
CLS 10\% & -1.511 & -0.624 & \textbf{-0.367} & -0.599 & -0.598 & -0.365\\
CLS 20\% & -1.512 & -0.624 & \textbf{-0.367} & -0.599 & -0.599 & -0.366\\
CLS 80\% & -1.515 & -0.624 & -0.370 & -0.602 & \textbf{-0.597} &
\textbf{-0.364}\\
CLS 90\% & -1.514 & -0.624 & -0.369 & -0.602 & \textbf{-0.597} &
\textbf{-0.364}\\\hline\hline
\end{tabular}
\end{table}

\begin{table}[ptb]
\caption{{\protect\footnotesize Average out-of-sample scores under a
misspecified Gaussian ARCH(1) model (Scenario (ii) in Table 1). All results
are based on }$\tau=5,000${\protect\footnotesize \ out-of-sample values.
Panels A, B and C, respectively, report the average scores when the true DGP
is GARCH(1,1) with }$t_{\nu=3},$ $t_{\nu=10}$ {\protect\footnotesize and}
$t_{\nu=30}${\protect\footnotesize \ errors. The rows in each panel refer to
the optimizer used. The columns refer to the out-of-sample measure. The
figures in bold are the largest average scores according to a given
out-of-sample measure.\medskip}}%
\label{tab:simgarchnu3}%
\centering\resizebox{12.3cm}{!}{
	\begin{tabular}{lrrrrrr}
		\hline\hline
		                   &                        &                          &                              &                              &                              &                              \\
		                   &                                                   \multicolumn{6}{c}{\textbf{Panel A: The true DGP is GARCH(1,1)-$t_{\nu=3}$}}                                                    \\
		                   &                        &                          &                              &                              &                              &                              \\
		                   &                                                               \multicolumn{6}{c}{\textbf{Out-of-sample score}}                                                                \\ \cline{2-7}
		                   &                        &                          &                              &                              &                              &                              \\
		                   & \multicolumn{1}{l}{LS} & \multicolumn{1}{l}{CRPS} & \multicolumn{1}{l}{CLS 10\%} & \multicolumn{1}{l}{CLS 20\%} & \multicolumn{1}{l}{CLS 80\%} & \multicolumn{1}{l}{CLS 90\%} \\
		\textbf{Optimizer} &                        &                          &                              &                              &                              &                              \\ \cline{1-1}
		                   &                        &                          &                              &                              &                              &                              \\
		LS                 &        \textbf{-2.335} &                   -1.248 &                       -0.568 &                       -0.873 &                       -0.892 &                       -0.574 \\
		CRPS               &                 -2.452 &          \textbf{-1.233} &                       -0.625 &                       -0.929 &                       -0.967 &                       -0.654 \\
		CLS 10\%            &                 -2.752 &                   -2.120 &              \textbf{-0.520} &                       -0.843 &                       -1.311 &                       -0.960 \\
		CLS 20\%            &                 -2.472 &                   -1.519 &                       -0.528 &              \textbf{-0.834} &                       -1.045 &                       -0.704 \\
		CLS 80\%            &                 -2.489 &                   -1.532 &                       -0.725 &                       -1.049 &              \textbf{-0.841} &                       -0.526 \\
		CLS 90\%            &                 -2.736 &                   -2.093 &                       -0.957 &                       -1.287 &                       -0.842 &              \textbf{-0.513} \\ \hline\hline
		                   &                        &                          &                              &                              &                              &                              \\
		                   &                                                   \multicolumn{6}{c}{\textbf{Panel B: The true DGP is GARCH(1,1)-$t_{\nu=10}$}}                                                   \\
		                   &                        &                          &                              &                              &                              &                              \\
		                   &                                                               \multicolumn{6}{c}{\textbf{Out-of-sample score}}                                                                \\ \cline{2-7}
		                   &                        &                          &                              &                              &                              &                              \\
		                   & \multicolumn{1}{l}{LS} & \multicolumn{1}{l}{CRPS} & \multicolumn{1}{l}{CLS 10\%} & \multicolumn{1}{l}{CLS 20\%} & \multicolumn{1}{l}{CLS 80\%} & \multicolumn{1}{l}{CLS 90\%} \\
		\textbf{Optimizer} &                        &                          &                              &                              &                              &                              \\ \cline{1-1}
		                   &                        &                          &                              &                              &                              &                              \\
		LS                 &        \textbf{-2.517} &                   -1.678 &                       -0.533 &                       -0.853 &                       -0.837 &                       -0.511 \\
		CRPS               &                 -2.525 &          \textbf{-1.677} &                       -0.538 &                       -0.857 &                       -0.841 &                       -0.516 \\
		CLS 10\%            &                 -2.563 &                   -1.754 &              \textbf{-0.530} &              \textbf{-0.850} &                       -0.882 &                       -0.546 \\
		CLS 20\%            &                 -2.534 &                   -1.706 &                       -0.531 &              \textbf{-0.850} &                       -0.854 &                       -0.523 \\
		CLS 80\%            &                 -2.532 &                   -1.704 &                       -0.543 &                       -0.868 &              \textbf{-0.834} &              \textbf{-0.508} \\
		CLS 90\%            &                 -2.555 &                   -1.738 &                       -0.561 &                       -0.888 &              \textbf{-0.834} &              \textbf{-0.508} \\ \hline\hline
		                   &                        &                          &                              &                              &                              &                              \\
		                   &                                                   \multicolumn{6}{c}{\textbf{Panel C: The true DGP is GARCH(1,1)-$t_{\nu=30}$}}                                                   \\
		                   &                        &                          &                              &                              &                              &                              \\
		                   &                                                               \multicolumn{6}{c}{\textbf{Out-of-sample score}}                                                                \\ \cline{2-7}
		                   &                        &                          &                              &                              &                              &                              \\
		                   & \multicolumn{1}{l}{LS} & \multicolumn{1}{l}{CRPS} & \multicolumn{1}{l}{CLS 10\%} & \multicolumn{1}{l}{CLS 20\%} & \multicolumn{1}{l}{CLS 80\%} & \multicolumn{1}{l}{CLS 90\%} \\
		\textbf{Optimizer} &                        &                          &                              &                              &                              &                              \\ \cline{1-1}
		                   &                        &                          &                              &                              &                              &                              \\
		LS                 &        \textbf{-2.532} &          \textbf{-1.727} &                       -0.525 &                       -0.824 &                       -0.817 &                       -0.501 \\
		CRPS               &                 -2.535 &          \textbf{-1.727} &                       -0.527 &                       -0.825 &                       -0.818 &                       -0.503 \\
		CLS 10\%            &                 -2.563 &                   -1.776 &              \textbf{-0.524} &                       -0.824 &                       -0.846 &                       -0.525 \\
		CLS 20\%            &                 -2.538 &                   -1.733 &                       -0.525 &              \textbf{-0.823} &                       -0.822 &                       -0.506 \\
		CLS 80\%            &                 -2.538 &                   -1.735 &                       -0.528 &                       -0.829 &              \textbf{-0.816} &                       -0.501 \\
		CLS 90\%            &                 -2.553 &                   -1.758 &                       -0.540 &                       -0.842 &                       -0.817 &              \textbf{-0.500} \\ \hline\hline
	\end{tabular}}\end{table}

\begin{table}[ptb]
\caption{{\protect\footnotesize Average out-of-sample scores under a
misspecified Gaussian ARCH(1) model with a fixed marginal mean (Scenario (iii)
in Table 1). All results are based on }$\tau=5,000$%
{\protect\footnotesize \ out-of-sample values. Panels A, B and C,
respectively, report the average scores when the true DGP is GARCH(1,1) with
}$t_{\nu=3},$ $t_{\nu=10}$ {\protect\footnotesize and} $t_{\nu=30}%
${\protect\footnotesize \ errors. The rows in each panel refer to the
optimizer used. The columns refer to the out-of-sample measure. The figures in
bold are the largest average scores according to a given out-of-sample
measure.\medskip}}%
\label{tab:simgarchnu3_fm}%
\centering\resizebox{12.3cm}{!}{
\begin{tabular}{lrrrrrr}
	\hline\hline
	                   &                        &                          &                              &                              &                              &                              \\
	                   &                                                   \multicolumn{6}{c}{\textbf{Panel A: The true DGP is GARCH(1,1)-$t_{\nu=3}$}}                                                    \\
	                   &                        &                          &                              &                              &                              &                              \\
	                   &                                                               \multicolumn{6}{c}{\textbf{Out-of-sample score}}                                                                \\ \cline{2-7}
	                   &                        &                          &                              &                              &                              &                              \\
	                   & \multicolumn{1}{l}{LS} & \multicolumn{1}{l}{CRPS} & \multicolumn{1}{l}{CLS 10\%} & \multicolumn{1}{l}{CLS 20\%} & \multicolumn{1}{l}{CLS 80\%} & \multicolumn{1}{l}{CLS 90\%} \\
	\textbf{Optimizer} &                        &                          &                              &                              &                              &                              \\ \cline{1-1}
	                   &                        &                          &                              &                              &                              &                              \\
	LS                 &                 -2.335 &                   -1.248 &                       -0.568 &                       -0.873 &                       -0.892 &                       -0.574 \\
	CRPS               &                 -2.451 &          \textbf{-1.233} &                       -0.626 &                       -0.929 &                       -0.966 &                       -0.652 \\
	CLS 10\%            &        \textbf{-2.329} &                   -1.257 &                       -0.565 &                       -0.871 &              \textbf{-0.883} &              \textbf{-0.565} \\
	CLS 20\%            &        \textbf{-2.329} &                   -1.257 &                       -0.565 &                       -0.871 &              \textbf{-0.883} &              \textbf{-0.565} \\
	CLS 80\%            &                 -2.335 &                   -1.257 &              \textbf{-0.564} &                       -0.870 &                       -0.888 &                       -0.571 \\
	CLS 90\%            &                 -2.334 &                   -1.257 &              \textbf{-0.564} &              \textbf{-0.869} &                       -0.888 &                       -0.570 \\ \hline\hline
	                   &                        &                          &                              &                              &                              &                              \\
	                   &                                                   \multicolumn{6}{c}{\textbf{Panel B: The true DGP is GARCH(1,1)-$t_{\nu=10}$}}                                                   \\
	                   &                        &                          &                              &                              &                              &                              \\
	                   &                                                               \multicolumn{6}{c}{\textbf{Out-of-sample score}}                                                                \\ \cline{2-7}
	                   &                        &                          &                              &                              &                              &                              \\
	                   & \multicolumn{1}{l}{LS} & \multicolumn{1}{l}{CRPS} & \multicolumn{1}{l}{CLS 10\%} & \multicolumn{1}{l}{CLS 20\%} & \multicolumn{1}{l}{CLS 80\%} & \multicolumn{1}{l}{CLS 90\%} \\
	\textbf{Optimizer} &                        &                          &                              &                              &                              &                              \\ \cline{1-1}
	                   &                        &                          &                              &                              &                              &                              \\
	LS                 &                 -2.517 &                   -1.678 &                       -0.533 &                       -0.852 &              \textbf{-0.837} &                       -0.511 \\
	CRPS               &                 -2.524 &          \textbf{-1.677} &                       -0.538 &                       -0.857 &                       -0.841 &                       -0.515 \\
	CLS 10\%            &                 -2.519 &                   -1.680 &                       -0.534 &                       -0.853 &                       -0.838 &                       -0.511 \\
	CLS 20\%            &                 -2.519 &                   -1.679 &                       -0.534 &                       -0.853 &                       -0.838 &                       -0.511 \\
	CLS 80\%            &        \textbf{-2.515} &                   -1.679 &              \textbf{-0.531} &              \textbf{-0.851} &              \textbf{-0.837} &              \textbf{-0.510} \\
	CLS 90\%            &        \textbf{-2.515} &                   -1.679 &              \textbf{-0.531} &              \textbf{-0.851} &              \textbf{-0.837} &              \textbf{-0.510} \\ \hline\hline
	                   &                        &                          &                              &                              &                              &                              \\
	                   &                                                   \multicolumn{6}{c}{\textbf{Panel C: The true DGP is GARCH(1,1)-$t_{\nu=30}$}}                                                   \\
	                   &                        &                          &                              &                              &                              &                              \\
	                   &                                                               \multicolumn{6}{c}{\textbf{Out-of-sample score}}                                                                \\ \cline{2-7}
	                   &                        &                          &                              &                              &                              &                              \\
	                   & \multicolumn{1}{l}{LS} & \multicolumn{1}{l}{CRPS} & \multicolumn{1}{l}{CLS 10\%} & \multicolumn{1}{l}{CLS 20\%} & \multicolumn{1}{l}{CLS 80\%} & \multicolumn{1}{l}{CLS 90\%} \\
	\textbf{Optimizer} &                        &                          &                              &                              &                              &                              \\ \cline{1-1}
	                   &                        &                          &                              &                              &                              &                              \\
	LS                 &        \textbf{-2.532} &          \textbf{-1.727} &              \textbf{-0.525} &              \textbf{-0.824} &              \textbf{-0.817} &              \textbf{-0.501} \\
	CRPS               &                 -2.535 &          \textbf{-1.727} &                       -0.527 &                       -0.825 &                       -0.818 &                       -0.502 \\
	CLS 10\%            &                 -2.535 &                   -1.728 &                       -0.526 &              \textbf{-0.824} &                       -0.819 &                       -0.503 \\
	CLS 20\%            &                 -2.535 &                   -1.728 &                       -0.526 &              \textbf{-0.824} &                       -0.818 &                       -0.503 \\
	CLS 80\%            &        \textbf{-2.532} &                   -1.728 &              \textbf{-0.525} &              \textbf{-0.824} &              \textbf{-0.817} &              \textbf{-0.501} \\
	CLS 90\%            &        \textbf{-2.532} &                   -1.728 &              \textbf{-0.525} &              \textbf{-0.824} &              \textbf{-0.817} &              \textbf{-0.501} \\ \hline\hline
\end{tabular}}\end{table}

\subsection{Visualization of strict coherence: score
densities\label{densities}}

We now provide further insights into the results in Tables
\ref{tab:simgarchTrueSpec}-\ref{tab:simgarchnu3_fm}, including the lack of
strict coherence in Table \ref{tab:simgarchnu3_fm}, by providing useful
approaches for visualizing strict coherence, and its absence.

Reiterating: under correct specification of the predictive model, in the limit
all predictives optimized according to criteria based on proper scoring rules
will yield equivalent predictive performance out-of-sample. In contrast, under
misspecification we expect that each score\textbf{\ }criterion will yield, in
principle, a distinct optimizing predictive and, hence, that\textbf{\ }%
out-of-sample performance will differ across predictives{; }with an optimizing
predictive expected to beat all others in terms of \textit{that} criterion.
Therefore, a lack of evidence in favour of strict coherence, in the presence
of misspecification, implies that the conjunction of the model and scoring
rule is unable to produce sufficiently distinct optimizers to, in turn, yield
distinct out-of-sample performance.

It is possible to shed light on this phenomenon by considering the limiting
behavior of the optimizers for the various scoring rules, across different
model specification regimes (reflecting those scenarios given in Table
\ref{tab:desing}). To this end, define $\boldsymbol{g}_{t}(\boldsymbol{\theta
}_{\ast})=\left.  \frac{\partial S\left(  P_{\boldsymbol{\theta}}^{t-1}%
,y_{t}\right)  }{\partial\boldsymbol{\theta}}\right\vert _{\boldsymbol{\theta
}=\boldsymbol{\theta}_{\ast}}$ and $\boldsymbol{h}_{t}(\boldsymbol{\theta
}_{\ast})=\left.  \frac{\partial^{2}S\left(  P_{\boldsymbol{\theta}}%
^{t-1},y_{t}\right)  }{\partial\boldsymbol{\theta}\partial\boldsymbol{\theta
}^{\prime}}\right\vert _{\boldsymbol{\theta}=\boldsymbol{\theta}_{\ast}}$, and
the limit quantities $\boldsymbol{J}(\boldsymbol{\theta}_{\ast})=\lim
_{T\rightarrow\infty}$Var$\left[  \frac{1}{\sqrt{T}}\sum\nolimits_{t=2}%
^{T}\boldsymbol{g}_{t}(\boldsymbol{\theta}_{\ast})\right]  $ and
$\boldsymbol{H}(\boldsymbol{\theta}_{\ast})=\lim_{T\rightarrow\infty}\frac
{1}{T}\sum\nolimits_{t=2}^{T}E\left[  \boldsymbol{h}_{t}(\boldsymbol{\theta
}_{\ast})\right]  ,$ where $\boldsymbol{\theta}_{\ast}$ denotes the maximum of
the limiting criterion function to which $\overline{S}(\boldsymbol{\theta})$
in (\ref{opt}) converges as $T$ diverges. Under regularity, the following
limiting distribution is in evidence: $\sqrt{T}(\widehat{\boldsymbol{\theta}%
}-\boldsymbol{\theta}_{\ast})\overset{d}{\rightarrow}N\left(  \boldsymbol{0}%
,\boldsymbol{V}_{\ast}\right)  ,$ where,
\begin{equation}
\boldsymbol{V}_{\ast}=\boldsymbol{H}^{-1}(\boldsymbol{\theta}_{\ast
})\boldsymbol{J}(\boldsymbol{\theta}_{\ast})\boldsymbol{H}^{-1}%
(\boldsymbol{\theta}_{\ast}). \label{vstar}%
\end{equation}

Under correct specification, and for criteria defined by proper scoring rules,
we expect that $\boldsymbol{\theta}_{\ast}=\boldsymbol{\theta}_{0}$ for all
versions of $\overline{S}(\boldsymbol{\theta})$. Given the efficiency of the
maximum likelihood estimator in this scenario, we would expect that the
sampling distribution of the optimizer associated with the log-score would be
more tightly concentrated around $\boldsymbol{\theta}_{\ast}$ than optimizers
associated with the other rules. However, since all optimizers would be
concentrating towards the same value, this difference would abate and
ultimately lead to scoring performances that are quite similar; i.e., a form
of strict coherence would not be in evidence, as is consistent with the
results in Table \ref{tab:simgarchTrueSpec}.

In contrast, under misspecification we expect that $\boldsymbol{\theta}_{\ast
}\neq\boldsymbol{\theta}_{0}$, with different optimizers consistent for
different values of\textbf{\ }$\boldsymbol{\theta}_{\ast}$. While the sampling
distributions of the different\textbf{\ }optimizers \textit{may} differ
substantially from each other, thereby leading to a form of strict coherence
as in Table \ref{tab:simgarchnu3}, this is not guaranteed to occur. Indeed, it
remains entirely possible that the resulting optimizers, while distinct, have
sampling distributions that are quite similar, even for very large values of
$T$.\footnote{This could occur for two (non-exclusive) reasons: one, the
variances $V_{\ast}$ in \eqref{vstar} are large; 2) the different limiting
optimized values are very similar. In either case, the sampling distributions
that result from this optimization procedure are likely to be very similar.}
In this case, the sampling distribution of the out-of-sample \textquotedblleft
optimized\textquotedblright\ $j$-th scoring rule $\overline{S}_{j}%
(\widehat{\boldsymbol{\theta}}^{[i]})$, evaluated at the $i$-th optimizer,
will not vary significantly with $i$, and\textbf{\ }strict coherence will
likely not be in evidence, even for large sample sizes, even though the model
is misspecified (and the limit optimizers unique).

This behavior can be illustrated graphically by simulating and analyzing (an
approximation to) the sampling distribution of $\overline{S}_{j}%
(\widehat{\boldsymbol{\theta}}^{[i]})$. We begin by generating $T=10,000$
observations from the three `true' DGPs in Table \ref{tab:desing}, and
producing predictions from the corresponding assumed predictive in each of the
three scenarios: (i) to (iii). Using the simulated observations, and for each
scenario, we compute $\widehat{\boldsymbol{\theta}}^{[i]}$ in (\ref{opt_theta}%
) by maximizing $\overline{S}_{i}(\boldsymbol{\theta}):=\frac{1}{T-1}%
\sum_{t=2}^{T}S_{i}\left(  P_{\boldsymbol{\theta}}^{t-1},y_{t}\right)  ,$ for
$S_{i}$, $i\in$ \{LS, CLS 10\%, CLS 20\%, CLS 80\%, CLS 90\%\}. Coherence can
then be visualized by constructing and analyzing the density of $s_{j}%
^{i}=\overline{S}_{j}(\widehat{\boldsymbol{\theta}}^{[i]})$, for $i,j\in$
\{LS, CLS 10\%, CLS 20\%, CLS 80\%, CLS 90\%\}, denoted here as $f(s_{j}^{i}%
)$. That is, we are interested in the density of the $jth$ sample score
criterion evaluated at the $ith$ optimizer, where $f(s_{j}^{j})$ denotes the
density of the $jth$ score evaluated at its own optimizer. To approximate this
density we first simulate $\{\widehat{\boldsymbol{\theta}}_{m}^{[i]}%
\}_{m=1}^{M}$ from the corresponding sampling distribution of
$\widehat{\boldsymbol{\theta}}^{[i]}$: $N(\widehat{\boldsymbol{\theta}}%
^{[i]},\widehat{\boldsymbol{V}}_{\ast}/T)$, where $\widehat{\boldsymbol{V}%
}_{\ast}$ is the usual finite sample estimator of $\boldsymbol{V}_{\ast}$ in
(\ref{vstar}).\footnote{In particular, $\widehat{\boldsymbol{V}}_{\ast}$ is
obtained as the sample estimator of $\boldsymbol{V}_{\ast}$, where
$\theta_{\ast}$ is replaced with $\widehat{\theta}^{[i]}$, for the $ith$
rule.} Given the simulated draws $\{\widehat{\boldsymbol{\theta}}_{m}%
^{[i]}\}_{m=1}^{M}$, we then compute\textbf{\ }$s_{j,m}^{i}=\overline{S}%
_{j}(\widehat{\boldsymbol{\theta}}_{m}^{[i]})$ for $m=1,\dots,M$. Under
coherence,\textbf{ }we do not expect any (estimated) density, $\widehat{f}%
(s_{j}^{i})$, for $i\neq j$, to be located to the right of the (estimated)
score-specific density, $\widehat{f}(s_{j}^{j})$ as, with positively-oriented
scores, this would reflect an inferior performance of the optimal predictive.
Under \textit{strict} coherence, we expect $\widehat{f}(s_{j}^{j})$ to lie to
the right of all other densities, and for there to be little overlap in
probability mass between $\widehat{f}(s_{j}^{j})$ and any other density.

\begin{figure}[ptb]
\centering
\begin{tabular}
[c]{ccc}%
\textbf{Panel A} & \textbf{Panel B} & \textbf{Panel C}\\
&  & \\
\includegraphics[scale= 0.9]{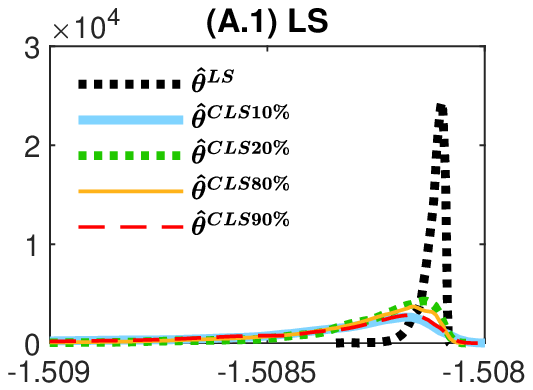} &
\includegraphics[scale= 0.9]{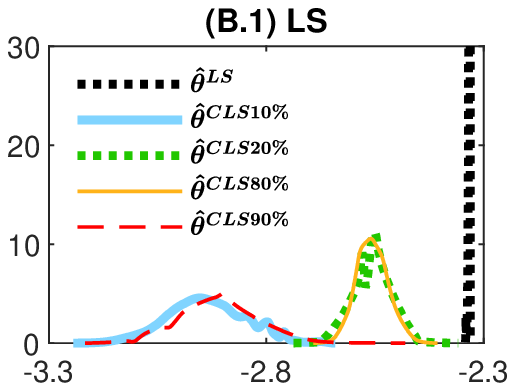} &
\includegraphics[scale= 0.9]{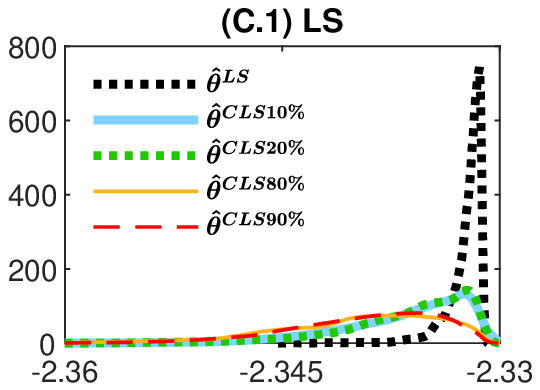}\\
&  & \\
\includegraphics[scale=0.9]{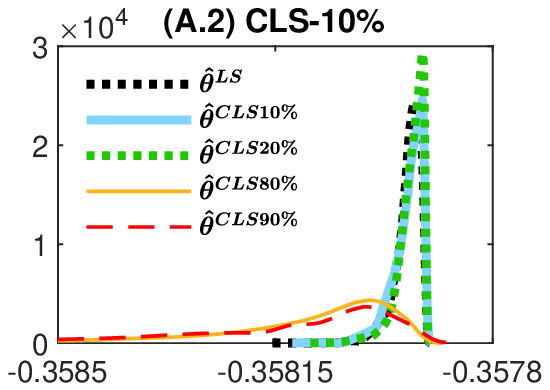} &
\includegraphics[scale=0.9]{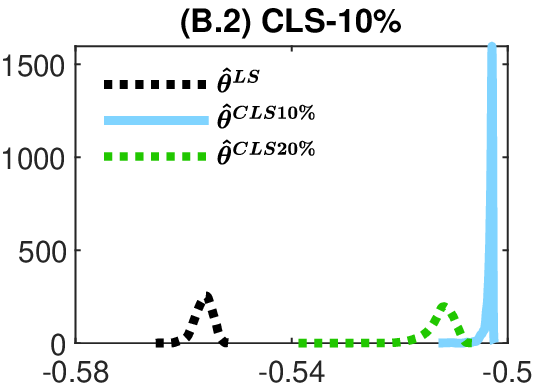} &
\includegraphics[scale=0.9]{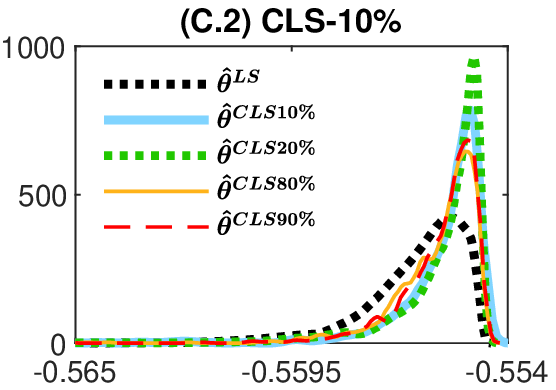}\\
&  & \\
\includegraphics[scale=0.9]{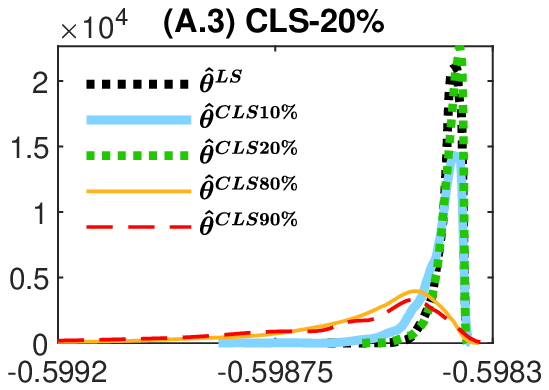} &
\includegraphics[scale=0.9]{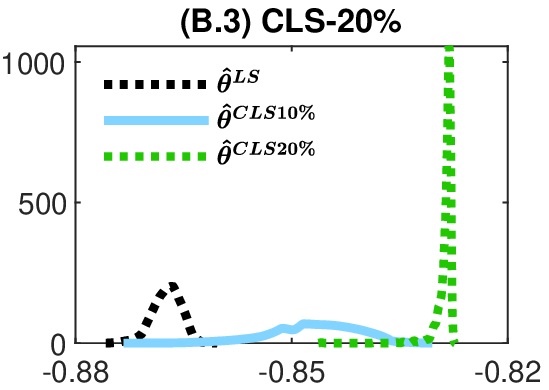} &
\includegraphics[scale=0.9]{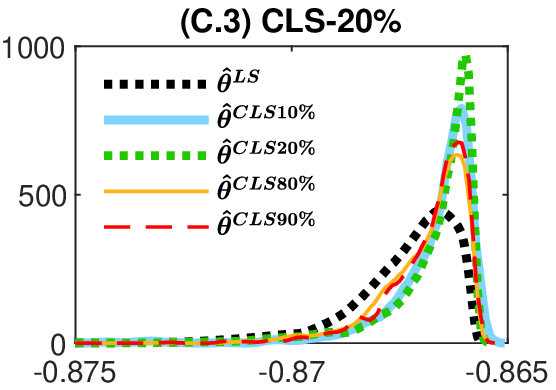}\\
&  &
\end{tabular}
\caption{{\protect\footnotesize Plots of the approximate density functions for
}$s_{j}^{i}=\overline{S}_{j}(\protect\widehat{\boldsymbol{\theta}}^{[i]}%
)${\protect\footnotesize . Panels A to C plot the density functions for }%
$j\in$ {\protect\footnotesize \{LS, CLS 10\%, CLS 20\%\}, evaluated at
}$\protect\widehat{\boldsymbol{\theta}}^{[i]}${\protect\footnotesize \ for
}$i\in$ {\protect\footnotesize \ \{LS, CLS 10\%, CLS 20\%, CLS 80\%, CLS
90\%\}. Panel A corresponds to the case of correct specification (Scenario (i)
in Table 1). Panel B corresponds to misspecification Scenario (ii) in Table 1.
Panel C corresponds to misspecification Scenario (iii) in Table 1. Each
sub-panel plots }$\protect\widehat{f}(s_{j}^{i})$ {\protect\footnotesize for
}$i\in${\protect\footnotesize \{LS, CLS 10\%, CLS 20\%, CLS 80\%, CLS 90\%\}
(as indicated in the key), and }$j\in${\protect\footnotesize \{LS, CLS 10\%,
CLS 20\%,\} (as indicated in the sub-panel heading). The reason for the
omission of results for }$i\in${\protect\footnotesize \{CLS 80\%, CLS 90\%\}
in sub-panels B.2 and B.3 is given in Footnote 4.}}%
\label{Fig:relativeaccuracy}%
\end{figure}

The results are given in Figure \ref{Fig:relativeaccuracy}. In the name of
brevity, we focus on Panels B and C of Figure \ref{Fig:relativeaccuracy},
which correspond respectively to Panels B and C of Table \ref{tab:desing}.
Each sub-panel in these two panels plots $\widehat{f}(s_{j}^{i})$ for $i\in$
\{LS, CLS 10\%, CLS 20\%, CLS 80\%, CLS 90\%\} (as indicated in the key), and
$j\in$ \{{LS, CLS 10\%, CLS 20\%,\} }(as indicated in the sub-panel heading).
The results in Panels B.1 to B.3 correspond to Scenario (ii) in Panel B of
Table \ref{tab:desing}. In this case, the impact of the misspecification is
stark. The score-specific ($i=j$) density in each case is far to the right of
the densities based on the other optimizers, and markedly more concentrated.
In Panels B.2 and B.3 we see that optimizing accordingly to \textit{some}%
\ sort of left-tail criterion, even if not that which matches the criterion
used to measure out-of-sample performance, produces densities that are further
to the right than those based on the log-score\textbf{\ }%
optimizer.\footnote{In these two sub-panels, the densities produced using
optimizers focussed on the tails that are opposite to those of predictive
interest (i.e. $\widehat{\boldsymbol{\theta}}^{[i]},$ $i\in$ \{CLS 80\%, CLS
90\%\}) are omitted, due to their being centred so far to the left of the
other densities, and being so dispersed, as to distort the figures.} In
contrast, we note in Panel B.1 that when the log-score itself is the
out-of-sample criterion of interest, it is preferable to use an optimizer that
focuses on a larger part of the support (either $\widehat{\boldsymbol{\theta}%
}^{[i]},$ $i\in$ \{CLS 20\%\} or $\widehat{\boldsymbol{\theta}}^{[i]},$ $i\in$
\{CLS 80\%\}), rather than one that focuses on the more extreme tails.
Moreover, due to the symmetry of both the true DGP and the assumed model, it
makes no difference (in terms of performance in log-score) which tail
optimizer (upper or lower) is used.

Panels C.1 to C.3\ correspond to Scenario (iii) in Panel C of Table
\ref{tab:desing}, with $\nu=3$ for the true DGP, and with predictions produced
via the misspecified Gaussian ARCH(1) model with the marginal mean fixed at
zero. The assumed\ model thus has no flexibility to shift location; this
feature clearly limiting the ability of the estimated predictive to assign
higher weight to the relevant part of the support when the realized
out-of-sample value demands it. As a consequence, there is no measurable gain
in using an optimizer that fits with the out-of-sample measure. These
observations are all consistent with the distinct similarity of all scores
(within a column) in Columns 1, 3 and 4 of Panel A in Table 4. In short:
strict coherence is not in evidence, despite the misspecification of the
predictive model. Just one simple, seemingly innocuous, change in
specification has been sufficient to eradicate the benefit of seeking an
optimal predictor. This suggests that even minor modifications to model
specification may\ have a significant impact in terms of the occurrence, or
otherwise, of strictly coherent predictions.

\subsection{Visualization of strict coherence: the role of sample
size\label{determ}}

The distinction between coherence and strict coherence can be couched in terms
of the distinction between the null hypothesis that two predictives - one
`optimal' and one not - have equal expected performance, and the alternative
hypothesis that the optimal predictive has superior expected performance. The
test of equal predictive ability of (any) two predictives{\ was a focus} of
\cite{giacomini2006tests} (GW hereafter; {see also related references:
\citealp{Diebold:1995}, \citealp{Hansen:2005}, and \citealp{Corradi:2006}});
hence, accessing the asymptotic distribution of their test statistic enables
us to shed some light on coherence. Specifically, what we do is solve the GW
test decision rule for the (out-of-sample) sample size required to yield
strict coherence, under misspecification. This enables us to gauge how large
the sample size must be to differentiate between an optimal and a non-optimal
prediction, in any particular misspecified scenario. In terms of the
illustration in the previous section, this is equivalent to gauging how large
the sample size needs to be to enable the relevant score-specific density in
each figure in Panels B and C of Figure \ref{Fig:relativeaccuracy} to lie to
the right of the others.

For $i\neq j$, define $\Delta_{t}^{ji}=S_{j}\left(  p(y_{t}|\mathcal{F}%
_{t-1},\widehat{\boldsymbol{\theta}}^{[j]}),y_{t}\right)  -S_{j}\left(
p(y_{t}|\mathcal{F}_{t-1},\widehat{\boldsymbol{\theta}}^{[i]}),y_{t}\right)  $
and $\overline{\Delta}_{\tau}^{ji}=\frac{1}{\tau}\sum_{t=T-\tau+1}^{T}%
\Delta_{t}^{ji},$ where the subscript $\tau$ is used to make explicit the
number of out-of-sample evaluations used to compute the difference in the two
average scores. The test of equal predictive ability is a test of
$H_{0}:\mathbb{E}[\Delta_{t}^{ji}|\mathcal{F}_{t-1}]=0$ versus $H_{1}%
:\mathbb{E}[\Delta_{t}^{ji}|\mathcal{F}_{t-1}]\neq0.$ Following GW, under
$H_{0}$ $Z_{\tau}=\tau(\overline{\Delta}_{\tau}^{ji})^{2}/var_{\tau}%
(\Delta_{t}^{ji})\overset{d}{\rightarrow}\chi_{(1)}^{2},$ where $var_{\tau
}(\Delta_{t}^{ji})$ denotes the sample variance of $\Delta_{t}^{ji}$ computed
over the evaluation period of size $\tau.$ Hence, at the $\alpha\times100\%$
significance level, the null will be rejected when, for given values of
$(\overline{\Delta}_{\tau}^{ji})^{2}$ and $var_{\tau}(\Delta_{t}^{ji})$,%
\begin{equation}
\tau>\frac{\chi_{(1)}^{2}(1-\alpha)\times var_{\tau}(\Delta_{t}^{ji}%
)}{(\overline{\Delta}_{\tau}^{ji})^{2}}=\tau^{\ast}, \label{tau}%
\end{equation}
where $\chi_{(1)}^{2}(1-\alpha)$ denotes the relevant critical value of the
limiting $\chi_{(1)}^{2}$ distribution of the test statistic.

The right-hand-side of the inequality in (\ref{tau}), from now on denoted by
$\tau^{\ast}$, indicates the minimum number of out-of-sample evaluations
associated with detection of a{\ significant difference }between $\overline
{S}_{j}(\hat{\boldsymbol{\theta}}^{[j]})$ and $\overline{S}_{j}(\hat
{\boldsymbol{\theta}}^{[i]})$. {For the purpose of this exercise, if
}$\overline{\Delta}_{\tau}^{ji}<0${, we set }$\tau^{\ast}=\tau${, as no value
of }$\tau^{\ast}${\ will induce rejection of the null hypothesis in favour of
strict coherence, which is the outcome we are interested in.} The value of
$\tau^{\ast}$ thus depends, for any given $\alpha$, on the relative magnitudes
of the sample quantities, $var_{\tau}(\Delta_{t}^{ji})$ and $(\overline
{\Delta}_{\tau}^{ji})^{2}.$ At a heuristic level, if $(\overline{\Delta}%
_{\tau}^{ji})^{2}$ and $var_{\tau}(\Delta_{t}^{ji})$ converge in probability
to constants $c_{1}$ and $c_{2}$, at rates that are some function of $\tau$,
then we are interested in plotting $\tau^{\ast}$ as a function of $\tau$, and
discerning when (if) $\tau^{\ast}$ begins to stabilize at a particular value.
It is \textit{this} value that then serves as a measure of the `ease' with
which strict coherence is in evidence in any particular example.

\begin{figure}[ptb]
\centering
\begin{tabular}
[c]{ccc}%
\includegraphics[scale= 0.7]{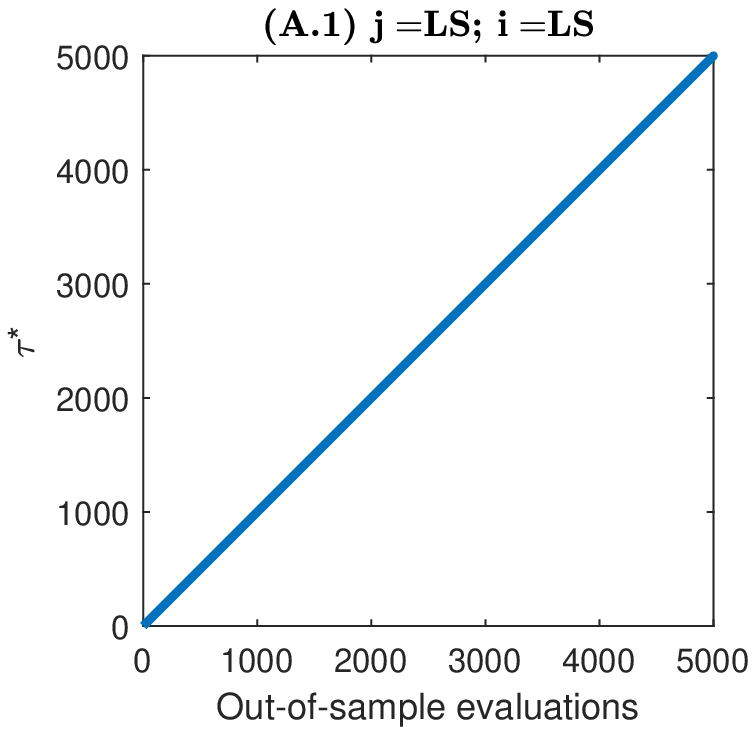} &
\includegraphics[scale=0.7]{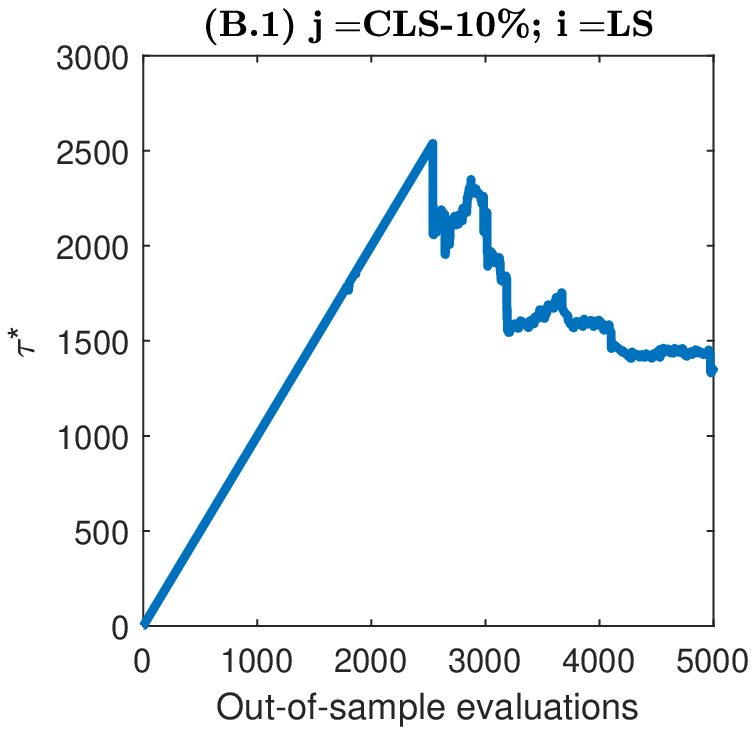} &
\includegraphics[scale=0.7]{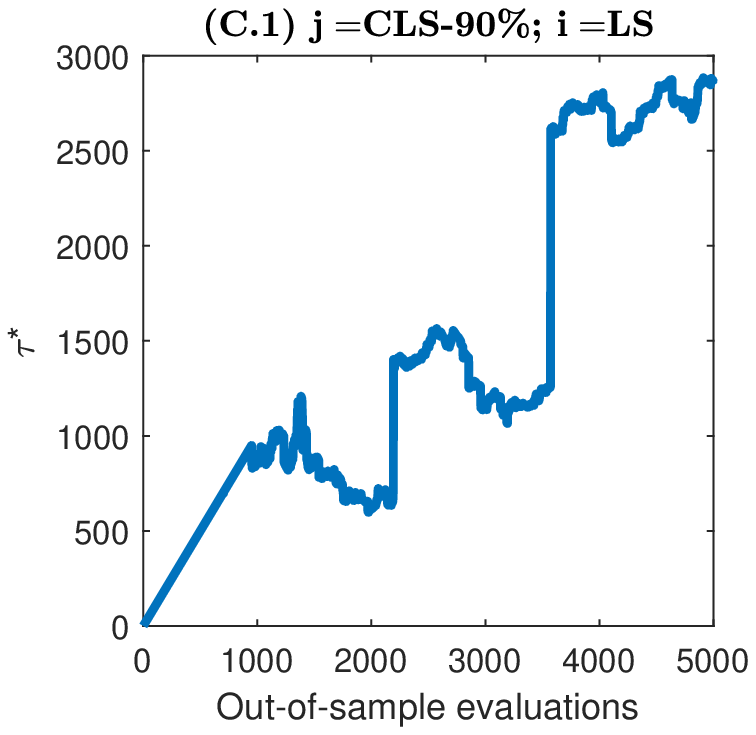}\\
&  & \\
\includegraphics[scale= 0.7]{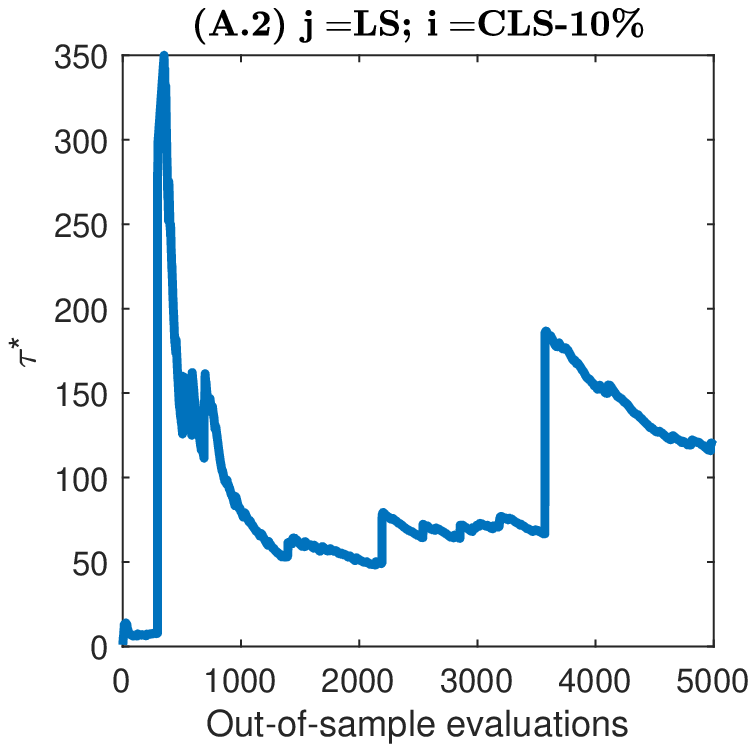} &
\includegraphics[scale=0.7]{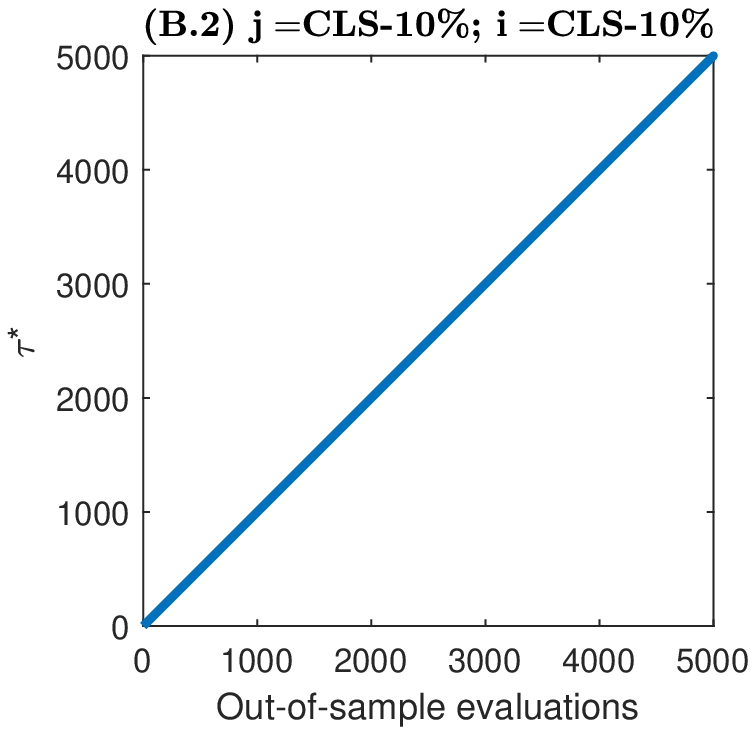} &
\includegraphics[scale=0.7]{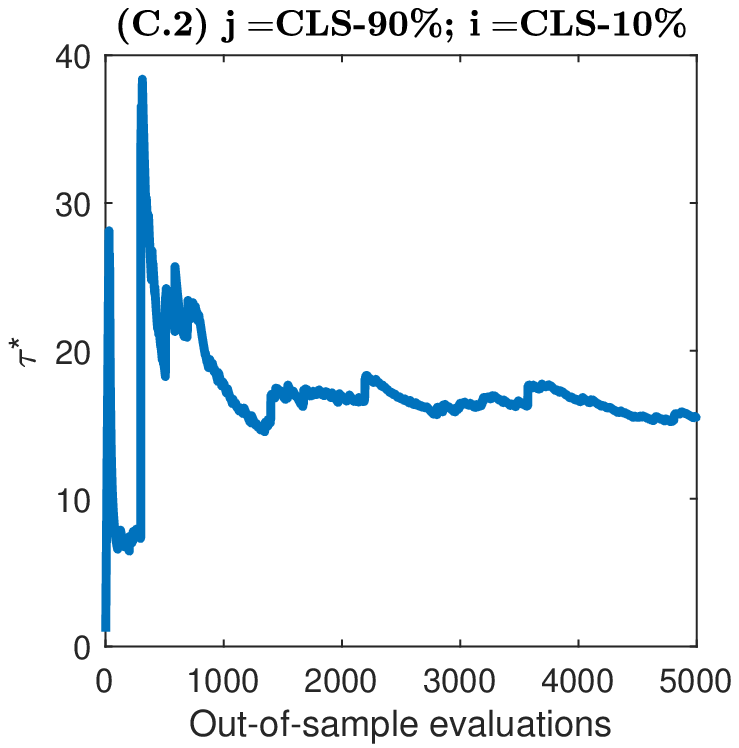}\\
&  & \\
\includegraphics[scale= 0.7]{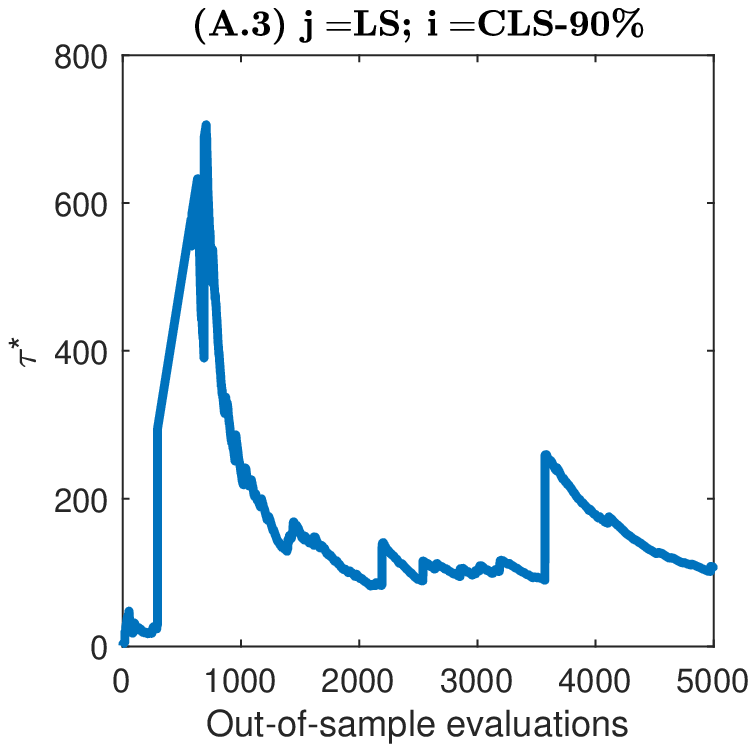} &
\includegraphics[scale=0.7]{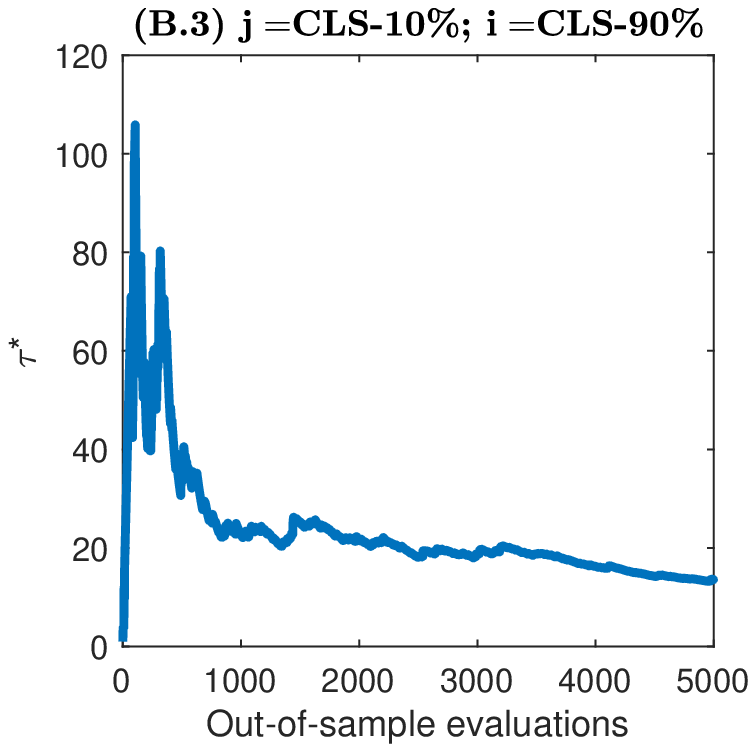} &
\includegraphics[scale=0.7]{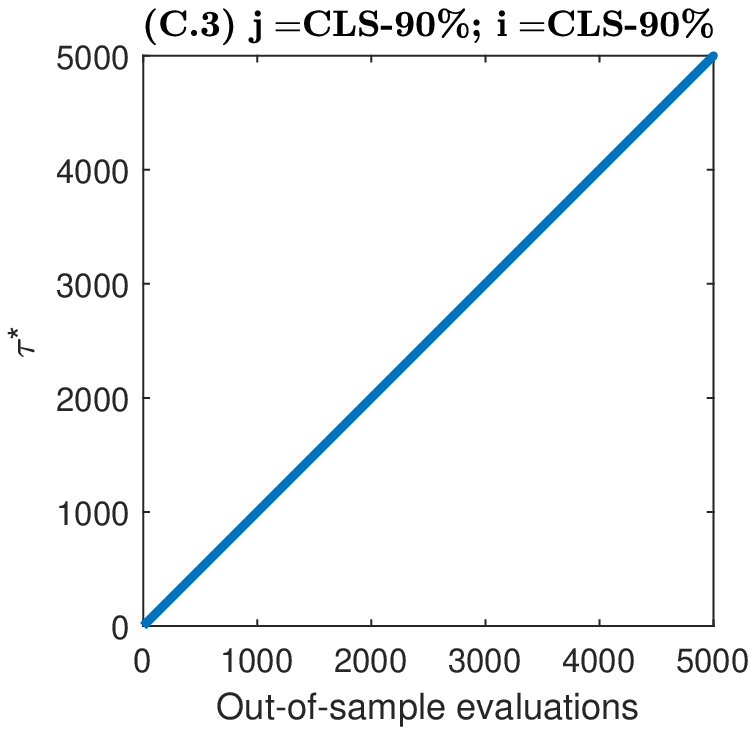}
\end{tabular}
\caption{{\protect\footnotesize Required number of out-of-sample evaluations
to reject }$H_{0}:\mathbb{E}[\Delta_{t}^{ji}|\mathcal{F}_{t-1}]=0$%
{\protect\footnotesize \ in favour of strict coherence: misspecification
Scenario (ii) in Table \ref{tab:desing}, where the true DGP is the
GARCH(1,1)-}$t_{\nu=3}$ {\protect\footnotesize model.}}%
\label{Fig:tau_starmissspec}%
\end{figure}

\begin{figure}[ptb]
\centering
\begin{tabular}
[c]{ccc}%
\includegraphics[scale= 0.7]{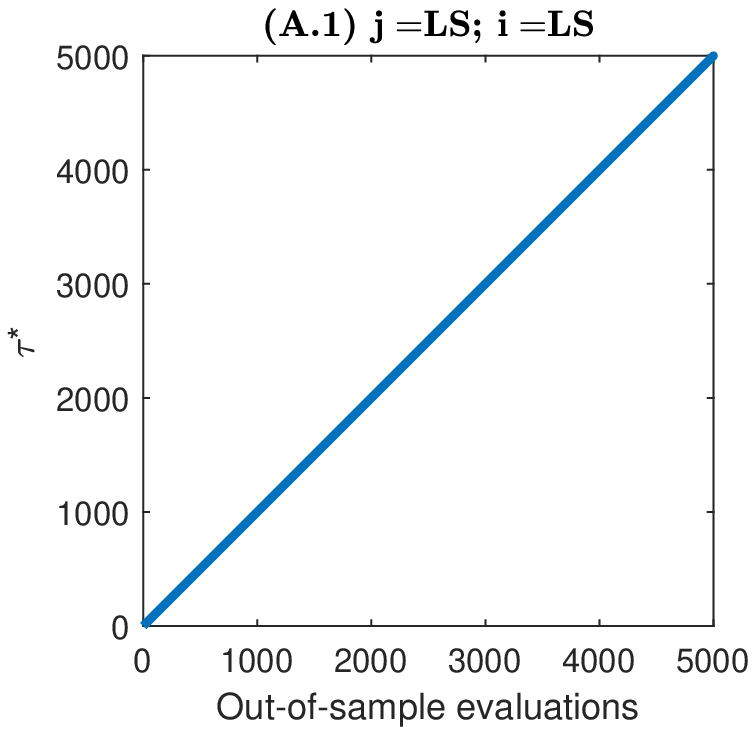} &
\includegraphics[scale=0.7]{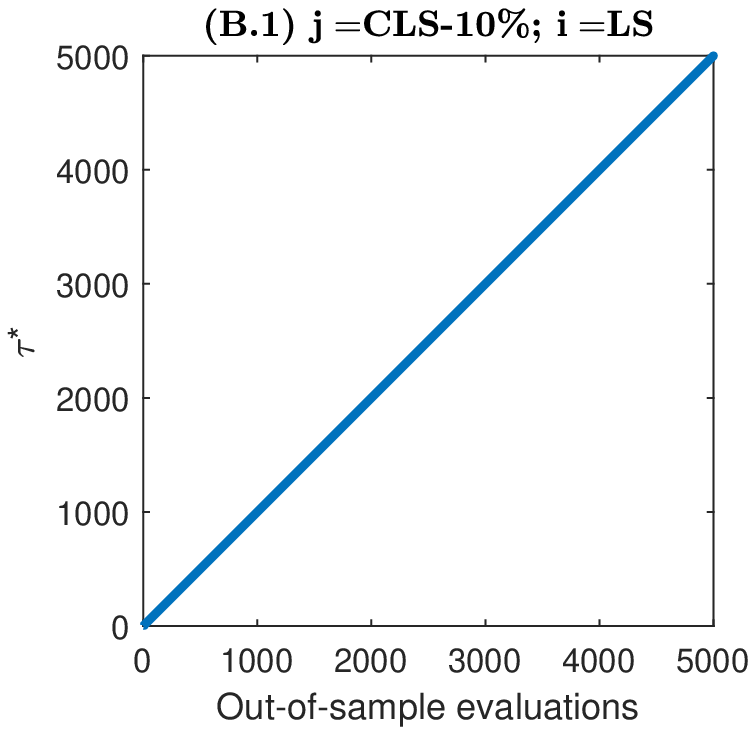} &
\includegraphics[scale=0.7]{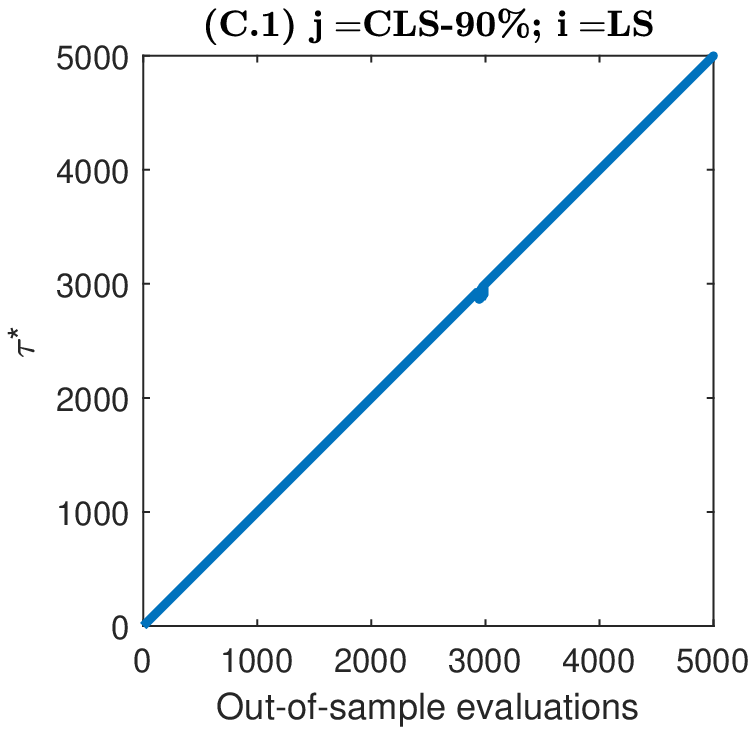}\\
&  & \\
\includegraphics[scale= 0.7]{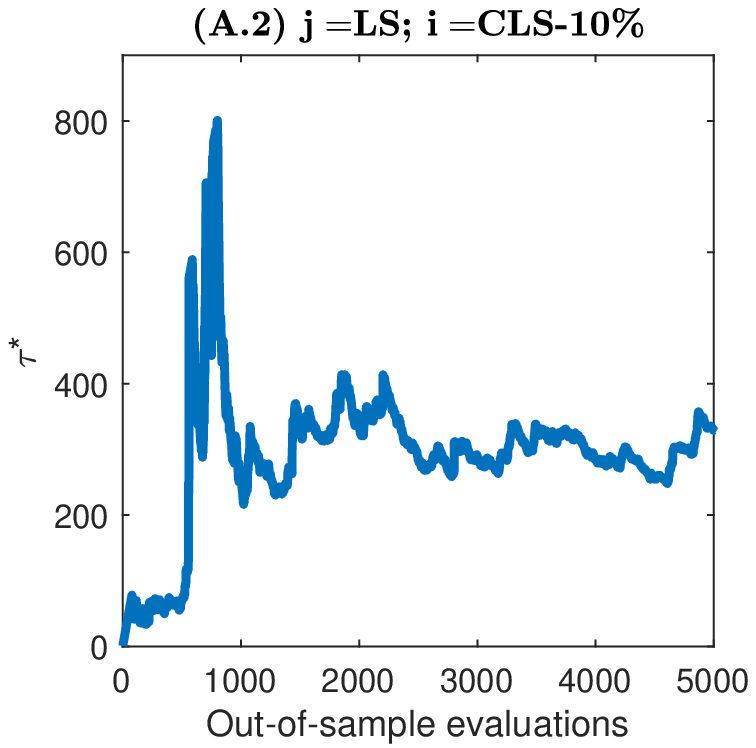} &
\includegraphics[scale=0.7]{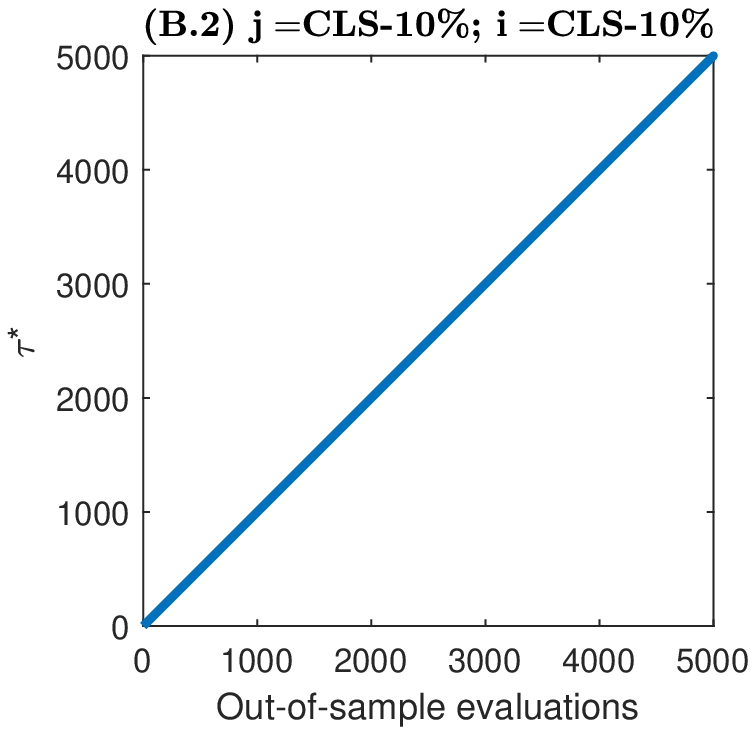} &
\includegraphics[scale=0.7]{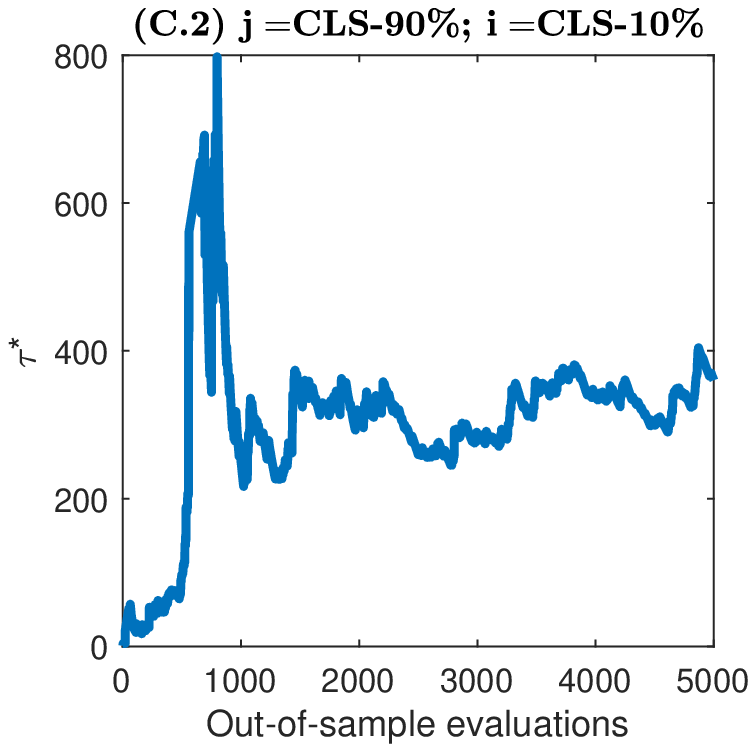}\\
&  & \\
\includegraphics[scale= 0.7]{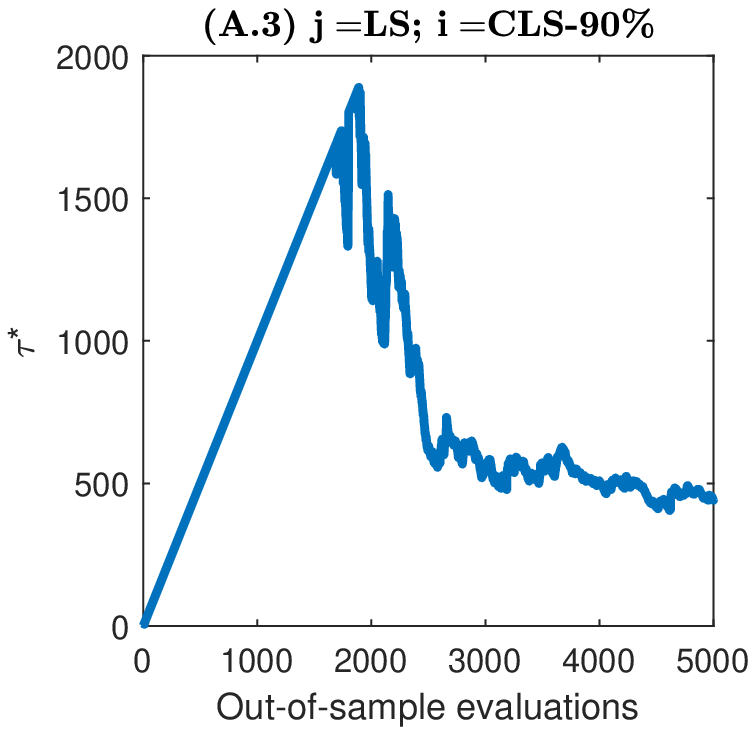} &
\includegraphics[scale=0.7]{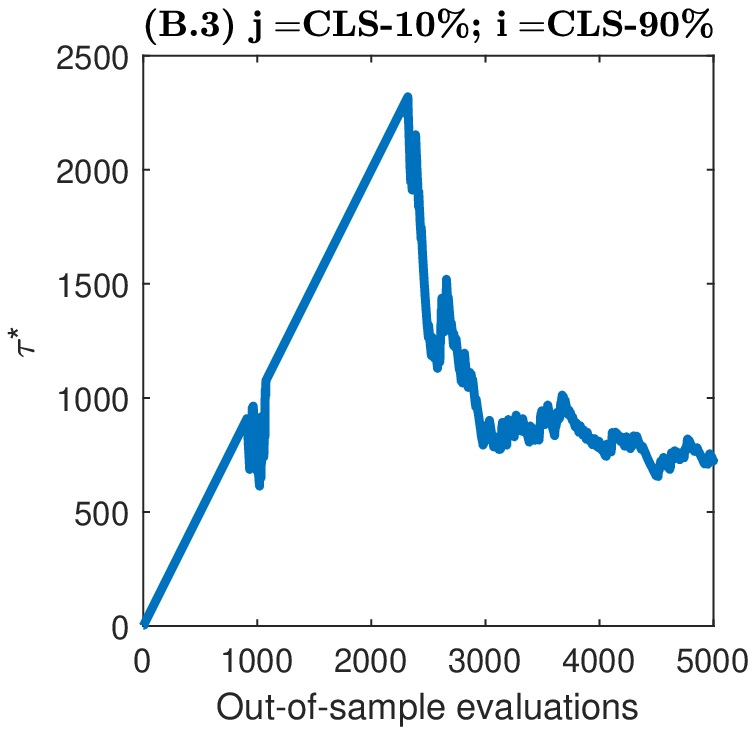} &
\includegraphics[scale=0.7]{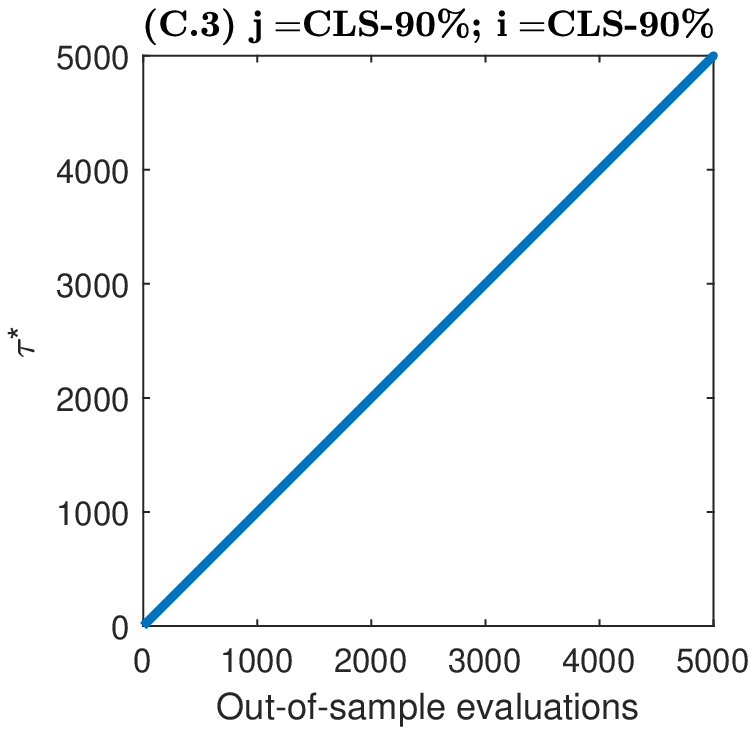}
\end{tabular}
\caption{{\protect\footnotesize Required number of out-of-sample evaluations
to reject }$H_{0}:\mathbb{E}[\Delta_{t}^{ji}|\mathcal{F}_{t-1}]=0$%
{\protect\footnotesize \ in favour of strict coherence: misspecification
Scenario (ii) in Table \ref{tab:desing}, where the true DGP is the
GARCH(1,1)-}$t_{\nu=30}$ {\protect\footnotesize model.}}%
\label{Fig:tau_starmissspecnu30}%
\end{figure}

\begin{figure}[ptb]
\centering
\begin{tabular}
[c]{ccc}%
\includegraphics[scale= 0.7]{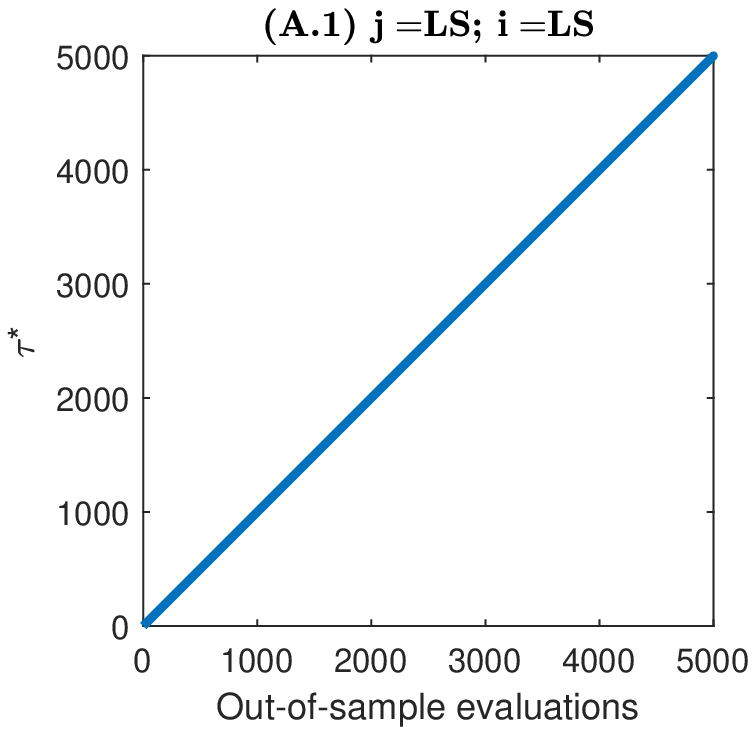} &
\includegraphics[scale=0.7]{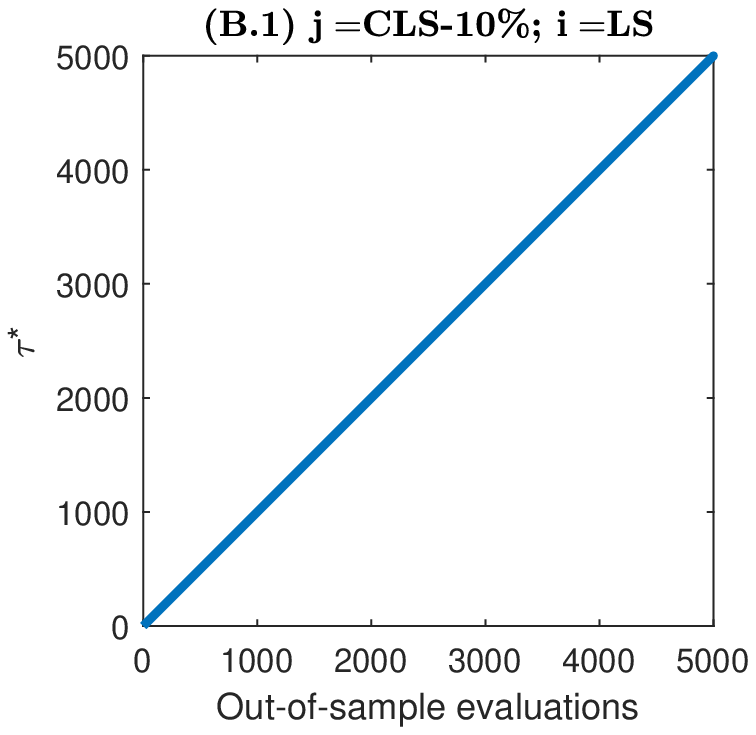} &
\includegraphics[scale=0.7]{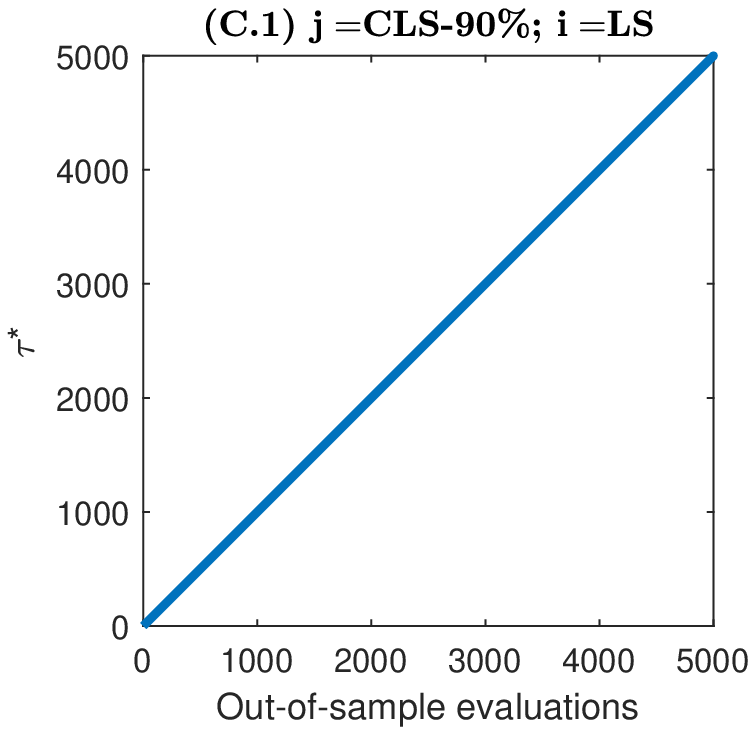}\\
&  & \\
\includegraphics[scale= 0.7]{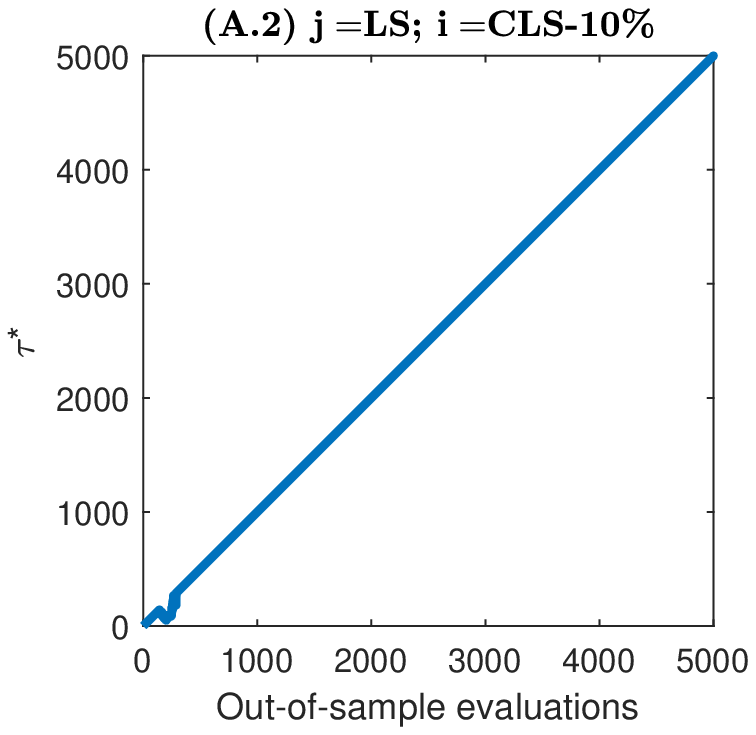} &
\includegraphics[scale=0.7]{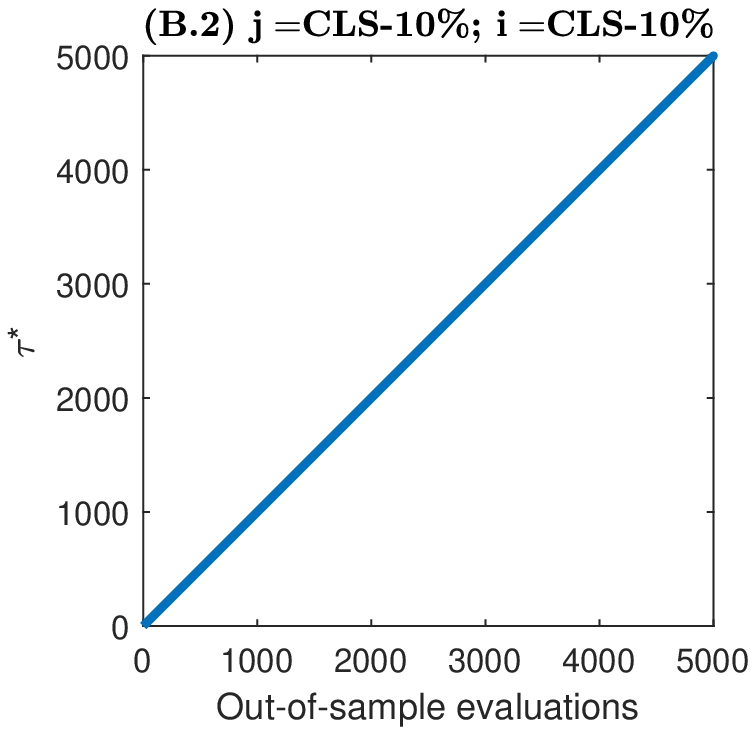} &
\includegraphics[scale=0.7]{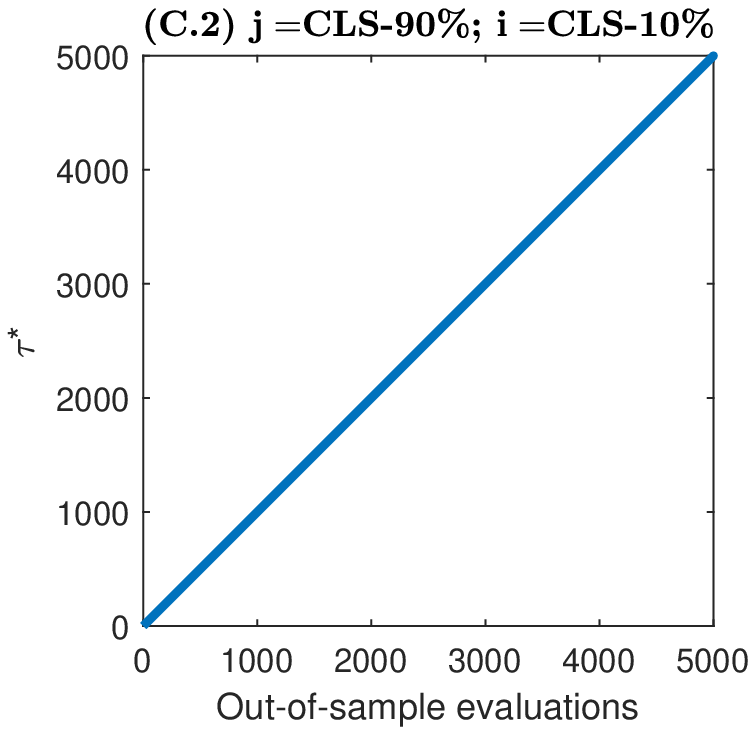}\\
&  & \\
\includegraphics[scale= 0.7]{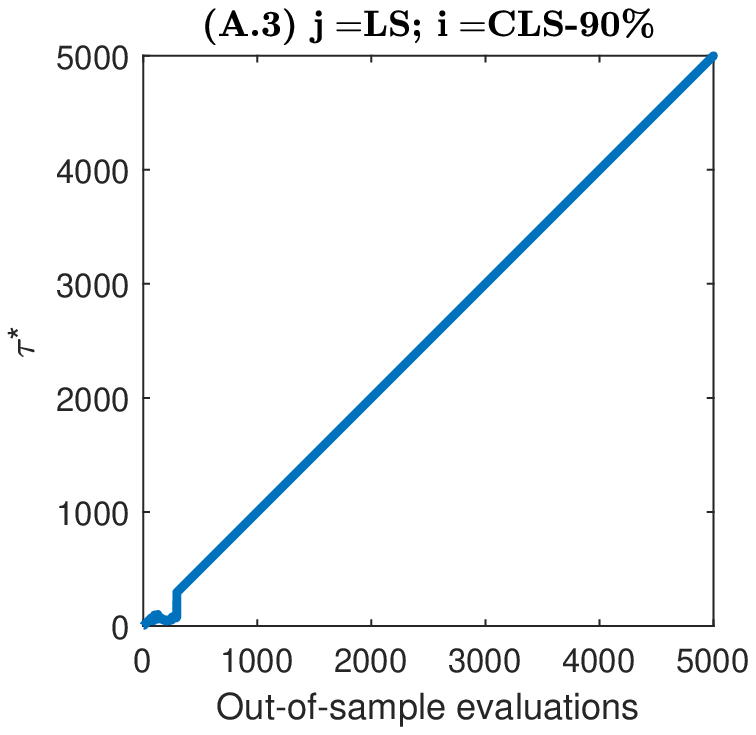} &
\includegraphics[scale=0.7]{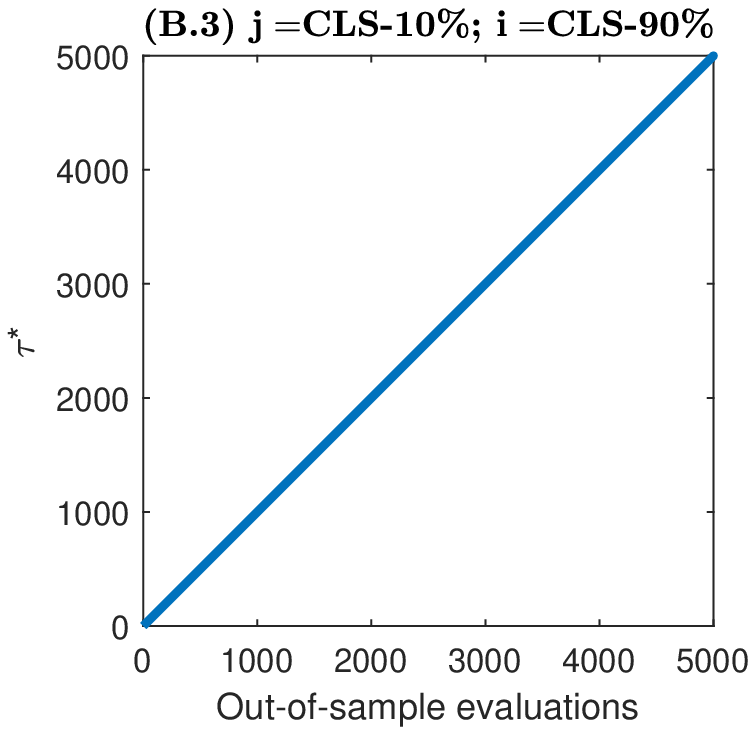} &
\includegraphics[scale=0.7]{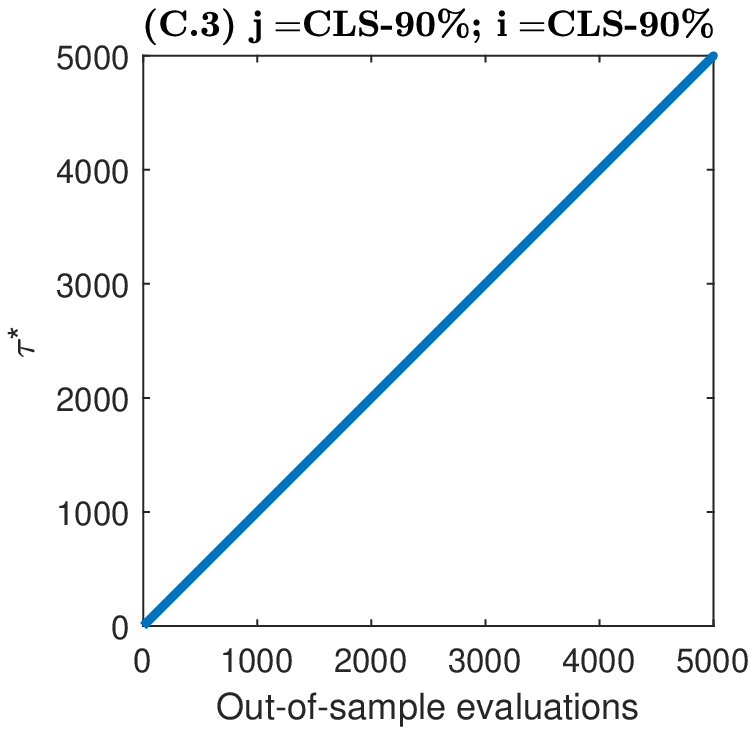}
\end{tabular}
\caption{{\protect\footnotesize Required number of out-of-sample evaluations
to reject }$H_{0}:\mathbb{E}[\Delta_{t}^{ji}|\mathcal{F}_{t-1}]=0$%
{\protect\footnotesize \ in favour of strict coherence: misspecification
Scenario (iii) in Table \ref{tab:desing}, where the true DGP is the
GARCH(1,1)-}$t_{\nu=3}$ {\protect\footnotesize model.}}%
\label{Fig:tau_star_misspecfixed}%
\end{figure}

In Figures 2 to 4 we plot $\tau^{\ast}$ as a function of $\tau$, for
$\tau=1,2,...,5,000$, and $\alpha=0.05$, for the misspecification scenarios
(ii) and (iii) in Table 1. In all figures, the diagonal panels simply plot a
45\% line, as these plots correspond to the case where $j=i$ and
$\overline{\Delta}_{\tau}^{ji}=0$ by construction. {Again, for} the purpose of
{the exercise} if $\overline{\Delta}_{\tau}^{ji}<0$, we set $\tau^{\ast}=\tau
$, as no value of $\tau^{\ast}$\ will induce {support of strict coherence}.
Moreover, whenever $\overline{\Delta}_{\tau}^{ji}>0$, but $\tau^{\ast}>\tau$,
we also set $\tau^{\ast}=\tau$. This allows us to avoid arbitrarily large
values of $\tau^{\ast}$\ that cannot be easily visualized. {These latter two
cases are thus also associated with 45\% lines. }%
Figures~\ref{Fig:tau_starmissspec} and \ref{Fig:tau_starmissspecnu30} report
results for Scenario (ii) with $\nu=3$ and $\nu=30$ respectively, whilst
Figure~\ref{Fig:tau_star_misspecfixed} presents the results for Scenario (iii)
with $\nu=3$. In each figure, sub-panels A.1 to A.3 record results for $j\in
$\ \{LS\}, and $i\in$\ \{LS, CLS 10\% and CLS 90\%\}. Sub-panels B.1 to B.3
record the corresponding results for $j\in$\ \{CLS 10\%\}, while sub-panels
C.1 to C.3 record the results for $j\in$\ \{CLS 90\%\}.

First consider sub-panels B.3 and C.2 in Figure \ref{Fig:tau_starmissspec}.
For $\tau>1,000$ (approximately), $\tau^{\ast}$ stabilizes at a value that is
approximately 20 in both cases. Viewing this value of $\tau^{\ast}$ as
`small', we conclude that it is `easy' to discern the strict coherence of an
upper tail optimizer relative to its lower tail counterpart, and vice versa,
under this form of misspecification. In contrast, Panels A.2 and A.3 indicate
that whilst strict coherence of the log-score optimizer is eventually
discernible, the value at which $\tau^{\ast}$ settles is larger (between about
100 and 200) than when the distinction is to be drawn between the two distinct
tail optimizers. Panels B.1 and C.1 show that it takes an even larger number
of out-of-sample observations ($\tau^{\ast}$\ exceeding $1,500$) to detect the
strict coherence of a tail optimizer relative to the log-score optimizer.
Indeed, from Panel C.1 it could be argued that the value of $\tau^{\ast}$
required to detect strict coherence relative to the log-score in the case of
CLS 90\% has not settled to a finite value even by $\tau=5,000.$

A comparison of Figures \ref{Fig:tau_starmissspec} and
\ref{Fig:tau_starmissspecnu30} highlights the effect of a reduction in
misspecification. In each off-diagonal sub-panel in Figure
\ref{Fig:tau_starmissspecnu30}, the value of $\tau^{\ast}$ is markedly higher
(i.e. more observations are required to detect strict coherence) than in the
corresponding sub-panel in Figure \ref{Fig:tau_starmissspec}. Indeed, Panel
C.1 in Figure \ref{Fig:tau_starmissspecnu30} indicates that strict coherence
in this particular case is, to all intents and purposes, unable to be
discerned in any reasonable number of out-of-sample observations. The
dissimilarity of the true DGP from the assumed model is simply not marked
enough for the optimal version of the CLS 90\% score to reap accuracy benefits
relative to the version of this score based on the log-score optimizer. This
particular scenario highlights the fact that, even if attainable, the pursuit
of coherence may not always be a practical endeavour. For example, if the
desired scoring rule is more computationally costly to evaluate than, say, the
log-score, then the small improvement in predictive accuracy yielded by
optimal prediction may not justify the added computational burden, in
particular for real-time forecasting exercises.

Finally, even more startling are the results in
Figure~\ref{Fig:tau_star_misspecfixed}, which we have termed the
`incompatible' case. For all out-of-sample scores considered, and all pairs of
optimizers, a diagonal line, $\tau^{\ast}=\tau$, results, as {either}
$\tau^{\ast}$ exceeds $\tau$ (and, hence, $\tau^{\ast}$ is set to $\tau$) for
all values of $\tau$, {or }$\overline{\Delta}_{\tau}^{ji}<0$, {in which case
}$\tau^{\ast}$ {is also set to} $\tau.$ Due to the incompatibility of the
assumed model with the true DGP strict coherence simply does not prevail in
any sense.

\subsection{Linear Pool Case: Simulation Design\label{linear_design}}

A common method of producing density forecasts from diverse models is to
consider the `optimal' combination of forecast (or predictive) densities
defined by a linear pool. Consider the setting where we entertain\textbf{\ }%
several possible models $\mathcal{M}_{k},$ $k=1,...,n$, all based on the same
information set, and with associated\textbf{\ }predictive distributions,
\begin{equation}
m_{k}(y_{t}|\mathcal{F}_{t-1}):=p(y_{t}|\mathcal{F}_{t-1},\boldsymbol{\theta
}_{k},\mathcal{M}_{k}),\text{ }k=1,...,n, \label{ms}%
\end{equation}
where the dependence of the $kth$ model on a $d_{k}$-dimensional set of
unknown parameters, $\boldsymbol{\theta}_{k}=(\theta_{k,1},\theta
_{k,2},...,\theta_{k,d_{k}})^{\prime}$, is captured in the short-hand
notation, $m_{k}(\cdot|\mathcal{\cdot}),$ and the manner in which
$\boldsymbol{\theta}_{k}$ is estimated is addressed below. The goal
is\textbf{\ }to determine how to combine the $n$ predictives in (\ref{ms}) to
produce an accurate forecast, in accordance with some measure of
predictive\ accuracy. As highlighted in the Introduction, we do not assume
that the true DGP coincides with any one of the constituent models in the
model set.

Herein, we follow \cite{mcconway1981marginalization}, and focus on the class
of linear combination processes only; i.e., the class of `linear pools' (see
also \citealp{genest1984pooling}, and \citealp{Geweke2011}):
\begin{equation}
\mathbb{P}:=\left\{  p(y_{t}|\mathcal{F}_{t-1},\mathbf{w}):=\sum_{k=1}%
^{n}w_{k}m_{k}(y_{t}|\mathcal{F}_{t-1});\quad\sum_{k=1}^{n}w_{k}=1;\quad\text{
and }\quad w_{k}\geq0\text{ }(k=1,...,n)\right\}  . \label{pred_class1}%
\end{equation}
Following the notion of optimal predictive estimation, and building on{\ }the
{established literature }cited earlier{, }we produce\ optimal weight
estimates
\begin{equation}
\hat{\mathbf{w}}:=\arg\max_{\mathbf{w}\in\Delta_{n}}\overline{S}%
(\mathbf{w}),\text{ where }\Delta_{n}:=\left\{  w_{k}\in\lbrack0,1]:\;\sum
_{k=1}^{n}w_{k}=1,\;w_{k}\geq0\;(k=1,...,n)\right\}  , \label{opt_w}%
\end{equation}
where\textbf{\ }$\overline{S}(\mathbf{w})$ is a sample average of the chosen
scoring rule, evaluated at the predictive distribution with density
$p(y_{t}|\mathcal{F}_{t-1},\mathbf{w})$, over a set of values defined below.
The estimator $\hat{\mathbf{w}}$ is referred to as the \textit{optimal score
estimator} (of $\mathbf{w}$) and the density $p(y_{t}|\mathcal{F}_{t-1}%
,\hat{\mathbf{w}})$ as the \textit{optimal linear pool. }The same set of
scoring rules as described in Section \ref{single_design} are adopted herein.

We simulate observations of $y_{t}$ from an autoregressive moving average
model of order (1,1) (ARMA(1,1)),
\begin{equation}
y_{t}=\phi_{1}+\phi_{2}y_{t-1}+\phi_{3}\varepsilon_{t-1}+\varepsilon_{t},
\label{true_dgp}%
\end{equation}
where $\phi_{1}=0$, $\phi_{2}=0.95$ and $\phi_{3}=-0.4.$ We employ five
different distributional assumptions for $\varepsilon_{t}$: $\varepsilon
_{t}\sim i.i.d.N\left(  0,1\right)  $, $\varepsilon_{t}\sim i.i.d.t_{v}\left(
0,1\right)  $, with $\nu=\left(  5,10,30\right)  $, and a mixture of normal
distributions $\varepsilon_{t}\sim i.i.d.\left[  pN\left(  \mu_{1},\sigma
_{1}\right)  +\left(  1-p\right)  N\left(  \mu_{2},\sigma_{2}\right)  \right]
$. In the case of the mixture, we set $\mu_{1}=0.3$, $\mu_{2}=-1.2$,
$\sigma_{1}=0.54$, $\sigma_{2}=1.43$ and $p=0.8$ to ensure that $E\left(
\varepsilon_{t}\right)  =0$ and $Var\left(  \varepsilon_{t}\right)  =1$, with
this setting inducing a negative skewness of -1.58. In constructing the model
pool, we consider three constituent models:
\begin{align}
\mathcal{M}_{1}:\text{ }y_{t}  &  \sim i.i.d.N(\theta_{1,1},\theta
_{1,2})\label{m1}\\
\mathcal{M}_{2}:\text{ }y_{t}  &  =\theta_{2,1}+\theta_{2,2}y_{t-1}+\eta
_{t}\text{ with }\eta_{t}\sim i.i.d.N(0,\theta_{2,3})\label{m2}\\
\mathcal{M}_{3}:\text{ }y_{t}  &  =\theta_{3,1}+\theta_{3,2}\eta_{t-1}%
+\eta_{t}\text{ with }\eta_{t}\sim i.i.d.N(0,\theta_{3,3}). \label{m3}%
\end{align}
All designs thus correspond to some degree of misspecification, with less
misspecification occurring when the true error term is either normal or
Student-t with a large value for $\nu$. Use of a skewed error term in
(\ref{true_dgp}) arguably produces the most extreme case of misspecification
and, hence, is the case where we would expect strict coherence to be most
evident.\footnote{Whilst there is no one model that corresponds to the true
DGP in (\ref{true_dgp}), an appropriately weighted sum of the three
predictives would be able to reproduce certain key features of the true
predictive, such as the autocorrelation structure, at least if the parameters
in each constituent model were set to appropriate values.}

For each design scenario, we take the following steps:

\begin{enumerate}
\item Generate $T$ observations of $y_{t}$ from the true DGP;

\item Use observations $t=1,...,J$, where $J=1,000$, to compute
$\widehat{\boldsymbol{\theta}}_{i}$ as in (\ref{opt_theta}) for each model
$m_{k}$, \newline for $S_{i}$, $i\in$ \{LS, CRPS, CLS 10\%, CLS 20\%, CLS 80\%
and CLS 90\%\};

\item For each $k=1,2,3$, construct the one-step-ahead predictive density
$\widehat{m}_{k}(y_{t}|\mathcal{F}_{t-1})=p(y_{t}|\mathcal{F}_{t-1}%
,\widehat{\boldsymbol{\theta}}_{i},\mathcal{M}_{k})$, for $t=J+1,...,J+\zeta$,
and compute $\widehat{\mathbf{w}}=(\widehat{w}_{1},\widehat{w}_{2}%
,\widehat{w}_{3})^{\prime}$ based on these $\zeta=50$ sets of predictive
densities as in (\ref{opt_w}), with $\overline{S}(\mathbf{w}):=\frac{1}{\zeta
}\sum_{t=J+1}^{J+\zeta}S\left(  P_{\widehat{\boldsymbol{\theta}},\mathbf{w}%
}^{t-1},y_{t}\right)  $, where $\widehat{\boldsymbol{\theta}}%
=(\widehat{\boldsymbol{\theta}}_{1},\widehat{\boldsymbol{\theta}}%
_{2},\widehat{\boldsymbol{\theta}}_{3})^{\prime}$ and
$P_{\widehat{\boldsymbol{\theta}},\mathbf{w}}^{t-1}$ is the predictive
distribution associated with the density $p(y_{t}|\mathcal{F}_{t-1}%
,\mathbf{w})$ in (\ref{pred_class1}).

\item Use $\widehat{\mathbf{w}}$ to obtain the pooled predictive density for
time point $t=J+\zeta+1$, $p(y_{t}|\mathcal{F}_{t-1},\widehat{\mathbf{w}%
})=\sum_{k=1}^{n=3}\widehat{w}_{k}\widehat{m}_{k}(y_{t}|\mathcal{F}_{t-1}).$

\item Roll the estimation sample forward\textbf{\ }by one observation and
repeat Steps 2 to 4, using the (non-subscripted) notation\textbf{\ }%
$\widehat{\boldsymbol{\theta}}=(\widehat{\boldsymbol{\theta}}_{1}%
,\widehat{\boldsymbol{\theta}}_{2},\widehat{\boldsymbol{\theta}}_{3})^{\prime
}$ for the estimator of $\boldsymbol{\theta}=(\boldsymbol{\theta}%
_{1},\boldsymbol{\theta}_{2},\boldsymbol{\theta}_{3})^{\prime}$ and
$\widehat{\mathbf{w}}$ for the estimator of $\mathbf{w}$ based on each rolling
sample of size $J+\zeta$.\textbf{\ }Produce $\tau=T-(J+$ $\zeta)$ pooled
predictive densities, and compute:%
\begin{equation}
\overline{S}(\widehat{\boldsymbol{\theta}},\widehat{\mathbf{w}})=\frac{1}%
{\tau}\sum_{t=T-\tau+1}^{T}S\left(  P_{\widehat{\boldsymbol{\theta}%
},\widehat{\mathbf{w}}}^{t-1},y_{t}\right)  . \label{s_linear_pool}%
\end{equation}
The results are tabulated and discussed in Section \ref{pool scores}. To keep
the notation manageable, we have not made explicit the fact that\textbf{\ }%
$\widehat{\boldsymbol{\theta}}$ and\textbf{\ }$\widehat{\mathbf{w}}$ are
produced by a given choice of score criterion, which may or may not match the
score used to construct (\ref{s_linear_pool}). The notation
$P_{\widehat{\boldsymbol{\theta}},\widehat{\mathbf{w}}}^{t-1}$\ refers to the
predictive distribution associated with the density $p(y_{t}|F_{t-1}%
,\widehat{\mathbf{w}}).$
\end{enumerate}

\subsection{Linear Pool Case: Simulation Results\label{pool scores}}

With reference to the results in Table \ref{tab:simnorm}, our expectations are
borne out to a large extent. The average out-of-sample scores in Panel B
pertain to arguably the most misspecified case, with the mixture of normals
inducing skewness in the true DGP, a feature that is not captured by any of
the components of the predictive pool. Whilst not uniformly indicative of
strict coherence, the results for this case\textbf{\ }are close to being so.
In particular, the optimal pools based on the CLS 20\%, CLS 80\% and CLS 90\%
criteria always {beat everything} else out of sample, according to each of
those same measures (i.e. the bold values appear on the diagonal in the last
three columns in Panel B). To two decimal places, the bold value also appears
on the diagonal in the column for the out-of-sample average of CLS 10\%. Thus,
the degree of misspecification of the model pool is sufficient to enable
strict coherence to be in evidence - most notably when it come to accurate
prediction in the tails. It can also be seen that log-score optimization reaps
benefits out-of-sample in terms of the log-score measure itself; only the CRPS
optimizer does not out-perform all others out-of-sample, in terms of the CRPS measure.

In contrast to the results in Panel B, those in Panel A (for the normal error
in the true DGP) are much more reminiscent of the `correct specification'
results in Table \ref{tab:simgarchTrueSpec}, in that all numbers within a
column are very similar, one to the other, and there is no marked diagonal
pattern. Interestingly however, given the earlier comments in the single model
context regarding the impact of the efficiency of the log-score optimizer
under correct specification, we note that the log-score optimizer yields the
smallest out-of-sample averages according to \textit{all }measures in Panel A.

This {superiority} of the log-score optimizer continues to be in evidence in
all three panels in Table \ref{tab:simt30}, in which the degrees of freedom in
the error term in the true DGP is successively increased, across the panels.
Moreover, there is arguably no more uniformity within columns in Panel C of
this table (in which the $t_{30}$ errors are a better match to the Gaussian
errors assumed in each component model in the pool), than there is in Panel A.
Clearly the use of the model pool is sufficient to pick up any degree of
fatness in the tails in the true DGP, so that no one design scenario is any
further from (or closer to) `correct specification' than the other. Hence,
what we observe in this table is simply a lack of strict coherence - i.e. the
degree of misspecification is not marked enough for score-specific optimizers
to reap benefits out-of-sample, and there is a good deal of similarity in the
performance of all optimizers, in any particular setting. Reiterating the
opening comment in this paragraph, in these settings of `near' to correct
specification, the efficiency of the log-score optimizer seems to be in
evidence. It is, in these cases, the only optimizer that one need to
entertain, no matter what the specific performance metric of interest!

\begin{table}[ptbh]
\caption{{\protect\footnotesize Average out-of-sample scores under two
different specifications for the true innovation, }$\varepsilon_{t}%
${\protect\footnotesize \ in (\ref{true_dgp}). Panel A (B) reports the average
scores based on }$\varepsilon_{t}\sim i.i.d.N(0,1)$ ($\varepsilon_{t}\sim
i.i.d.Mixture$ $of$ $normals$). {\protect\footnotesize The rows in each panel
refer to the optimizer used. The columns refer to the out-of-sample measure
used to compute the average scores. The figures in bold are the largest
average scores according to a given out-of-sample measure. All results are
based on }$\tau=5,000${\protect\footnotesize \ out-of-sample values. \medskip
}}%
\label{tab:simnorm}%
\centering%
\begin{tabular}
[c]{lrrrrrr}\hline\hline
&  &  &  &  &  & \\
& \multicolumn{6}{c}{\textbf{Panel A: $\varepsilon_{t}\sim i.i.d.N(0,1)$}}\\
&  &  &  &  &  & \\
& \multicolumn{6}{c}{\textbf{Out-of-sample score}}\\\cline{2-7}
&  &  &  &  &  & \\
& \multicolumn{1}{l}{LS} & \multicolumn{1}{l}{CRPS} & \multicolumn{1}{l}{CLS
10\%} & \multicolumn{1}{l}{CLS 20\%} & \multicolumn{1}{l}{CLS 80\%} &
\multicolumn{1}{l}{CLS 90\%}\\
\textbf{Optimizer} &  &  &  &  &  & \\\cline{1-1}
&  &  &  &  &  & \\
LS & \textbf{-1.493} & \textbf{-0.500} & \textbf{-0.250} & \textbf{-0.451} &
\textbf{-0.447} & \textbf{-0.251}\\
CRPS & -1.512 & -0.529 & -0.255 & -0.461 & -0.452 & -0.253\\
CLS 10\% & -1.709 & -0.570 & -0.254 & -0.465 & -0.608 & -0.376\\
CLS 20\% & -1.514 & -0.507 & -0.252 & -0.455 & -0.460 & -0.261\\
CLS 80\% & -1.532 & -0.518 & -0.271 & -0.478 & -0.451 & -0.253\\
CLS 90\% & -1.648 & -0.551 & -0.318 & -0.555 & -0.459 & -0.257\\\hline\hline
&  &  &  &  &  & \\
& \multicolumn{6}{c}{\textbf{Panel B:} $\varepsilon_{t}\sim i.i.d.Mixture$
$of$ $normals$}\\
&  &  &  &  &  & \\
& \multicolumn{6}{c}{\textbf{Out-of-sample score}}\\\cline{2-7}
&  &  &  &  &  & \\
& \multicolumn{1}{l}{LS} & \multicolumn{1}{l}{CRPS} & \multicolumn{1}{l}{CLS
10\%} & \multicolumn{1}{l}{CLS 20\%} & \multicolumn{1}{l}{CLS 80\%} &
\multicolumn{1}{l}{CLS 90\%}\\
\textbf{Optimizer} &  &  &  &  &  & \\\cline{1-1}
&  &  &  &  &  & \\
LS & \textbf{-1.479} & \textbf{-0.472} & -0.313 & -0.522 & -0.374 & -0.207\\
CRPS & -1.502 & -0.528 & -0.348 & -0.563 & -0.363 & -0.198\\
CLS10\% & -1.703 & -0.585 & -0.300 & -0.525 & -0.529 & -0.319\\
CLS20\% & -1.605 & -0.519 & \textbf{-0.297} & \textbf{-0.511} & -0.466 &
-0.275\\
CLS80\% & -1.772 & -0.494 & -0.557 & -0.824 & \textbf{-0.347} &
\textbf{-0.191}\\
CLS90\% & -2.319 & -0.580 & -0.863 & -1.246 & -0.358 & \textbf{-0.191}%
\\\hline\hline
\end{tabular}
\end{table}

\begin{table}[ptbh]
\caption{{\protect\footnotesize Average out-of-sample scores under two
different specifications for the true innovation, }$\varepsilon_{t}%
${\protect\footnotesize \ in (\ref{true_dgp}) Panel A (B; C) reports the
average scores based on }$\varepsilon_{t}\sim i.i.d.t_{\nu=5}$ ($\varepsilon
_{t}\sim i.i.d.t_{\nu=10};$ $\varepsilon_{t}\sim i.i.d.t_{\nu=30}%
$){\protect\footnotesize . The rows in each panel refer to the optimizer used.
The columns refer to the out-of-sample measure used to compute the average
scores. The figures in bold are the largest average scores according to a
given out-of-sample measure. All results are based on }$\tau=5,000$%
{\protect\footnotesize \ out-of-sample values. \medskip}}%
\label{tab:simt30}%
\centering\resizebox{14.3cm}{!}{
\begin{tabular}{lrrrrrr}
	\hline\hline
	                   &                        &                          &                              &                              &                              &                              \\
	                   &                                            \multicolumn{6}{c}{\textbf{Panel A: $\protect\varepsilon_t \sim i.i.d.t_{\nu=5}$}}                                             \\
	                   &                        &                          &                              &                              &                              &                              \\
	                   &                                                               \multicolumn{6}{c}{\textbf{Out-of-sample score}}                                                                \\ \cline{2-7}
	                   &                        &                          &                              &                              &                              &                              \\
	                   & \multicolumn{1}{l}{LS} & \multicolumn{1}{l}{CRPS} & \multicolumn{1}{l}{CLS 10\%} & \multicolumn{1}{l}{CLS 20\%} & \multicolumn{1}{l}{CLS 80\%} & \multicolumn{1}{l}{CLS 90\%} \\
	\textbf{Optimizer} &                        &                          &                              &                              &                              &                              \\ \cline{1-1}
	                   &                        &                          &                              &                              &                              &                              \\
	LS                 &        \textbf{-1.756} &          \textbf{-0.630} &              \textbf{-0.311} &              \textbf{-0.522} &              \textbf{-0.508} &              \textbf{-0.298} \\
	CRPS               &                 -1.782 &                   -0.676 &                       -0.317 &                       -0.534 &                       -0.515 &                       -0.304 \\
	CLS 10\%            &                 -1.881 &                   -0.706 &                       -0.315 &                       -0.540 &                       -0.591 &                       -0.358 \\
	CLS 20\%            &                 -1.805 &                   -0.656 &                       -0.312 &                       -0.528 &                       -0.545 &                       -0.326 \\
	CLS 80\%            &                 -1.810 &                   -0.665 &                       -0.339 &                       -0.560 &                       -0.514 &                       -0.300 \\
	CLS 90\%            &                 -1.909 &                   -0.732 &                       -0.375 &                       -0.617 &                       -0.532 &                       -0.306 \\ \hline\hline
	                   &                        &                          &                              &                              &                              &                              \\
	                   &                                            \multicolumn{6}{c}{\textbf{Panel B: $\protect\varepsilon_t \sim i.i.d.t_{\nu=10}$}}                                            \\
	                   &                        &                          &                              &                              &                              &                              \\
	                   &                                                               \multicolumn{6}{c}{\textbf{Out-of-sample score}}                                                                \\ \cline{2-7}
	                   &                        &                          &                              &                              &                              &                              \\
	                   & \multicolumn{1}{l}{LS} & \multicolumn{1}{l}{CRPS} & \multicolumn{1}{l}{CLS 10\%} & \multicolumn{1}{l}{CLS 20\%} & \multicolumn{1}{l}{CLS 80\%} & \multicolumn{1}{l}{CLS 90\%} \\
	\textbf{Optimizer} &                        &                          &                              &                              &                              &                              \\ \cline{1-1}
	                   &                        &                          &                              &                              &                              &                              \\
	LS                 &        \textbf{-1.611} &          \textbf{-0.557} &              \textbf{-0.263} &              \textbf{-0.478} &              \textbf{-0.475} &              \textbf{-0.274} \\
	CRPS               &                 -1.622 &                   -0.588 &                       -0.266 &                       -0.483 &                       -0.478 &                       -0.276 \\
	CLS 10\%            &                 -1.822 &                   -0.627 &                       -0.269 &                       -0.489 &                       -0.653 &                       -0.426 \\
	CLS 20\%            &                 -1.674 &                   -0.582 &                       -0.264 &                       -0.481 &                       -0.529 &                       -0.320 \\
	CLS 80\%            &                 -1.659 &                   -0.582 &                       -0.286 &                       -0.513 &                       -0.480 &                       -0.275 \\
	CLS 90\%            &                 -1.757 &                   -0.634 &                       -0.335 &                       -0.583 &                       -0.489 &                       -0.278 \\ \hline\hline
	                   &                        &                          &                              &                              &                              &                              \\
	                   &                                            \multicolumn{6}{c}{\textbf{Panel C: $\protect\varepsilon_t \sim i.i.d.t_{\nu=30}$}}                                            \\
	                   &                        &                          &                              &                              &                              &                              \\
	                   &                                                               \multicolumn{6}{c}{\textbf{Out-of-sample score}}                                                                \\ \cline{2-7}
	                   &                        &                          &                              &                              &                              &                              \\
	                   & \multicolumn{1}{l}{LS} & \multicolumn{1}{l}{CRPS} & \multicolumn{1}{l}{CLS 10\%} & \multicolumn{1}{l}{CLS 20\%} & \multicolumn{1}{l}{CLS 80\%} & \multicolumn{1}{l}{CLS 90\%} \\
	\textbf{Optimizer} &                        &                          &                              &                              &                              &                              \\ \cline{1-1}
	                   &                        &                          &                              &                              &                              &                              \\
	LS                 &        \textbf{-1.532} &          \textbf{-0.517} &              \textbf{-0.260} &              \textbf{-0.473} &              \textbf{-0.450} &              \textbf{-0.255} \\
	CRPS               &                 -1.553 &                   -0.547 &                       -0.265 &                       -0.483 &                       -0.457 &                       -0.258 \\
	CLS 10\%            &                 -1.909 &                   -0.619 &                       -0.264 &                       -0.487 &                       -0.747 &                       -0.484 \\
	CLS 20\%            &                 -1.559 &                   -0.530 &                       -0.262 &                       -0.477 &                       -0.470 &                       -0.271 \\
	CLS 80\%            &                 -1.572 &                   -0.538 &                       -0.285 &                       -0.505 &                       -0.453 &                       -0.256 \\
	CLS 90\%            &                 -1.748 &                   -0.588 &                       -0.368 &                       -0.623 &                       -0.463 &                       -0.261 \\ \hline\hline
\end{tabular}}\end{table}

\section{Empirical Illustration: Financial Returns\label{emp}}

\subsection{Overview}

We now illustrate the performance of optimal prediction in a realistic
empirical setting. We return to the earlier example of financial returns, but
with a range of increasingly sophisticated models used to capture the features
of observed data. Both single models and a linear pool are entertained. We
consider returns on two indexes:\ S\&P500 and MSCI Emerging Market (MSCIEM).
The data for both series extend from January 3rd, 2000 to May 7th, 2020. All
returns are continuously compounded \ in daily percentage units. For each time
series, we reserve the first 1,500 observations for the initial parameter
estimation, and conduct the predictive evaluation exercise for the period
between March 16th, 2006 and May 7th, 2020, with the predictive evaluation
period covering both the global financial crisis (GFC) and the recent downturn
caused by the COVID19 pandemic. As is consistent with the typical features
exhibited by financial returns, the descriptive statistics reported in Table
\ref{tab:summaries} provide evidence of time-varying and autocorrelated
volatility (significant serial correlation in squared returns) and marginal
non-Gaussianity (significant non-normality in the level of returns) in both
series, with evidence of slightly more negative skewness in the MSCIEM series.

\begin{table}[ptb]
\centering
\scalebox{0.85}{
	\begin{tabular}{lcccccccccc}
		\hline\hline
		        &         &        &        &        &        &        &          &          &         &         \\
		        &     Min &    Max &   Mean & Median & St.Dev &  Range & Skewness & Kurtosis & JB stat & LB stat (10) \\
		        &         &        &        &        &        &        &          &          &         &         \\
		S\&P500 & -12.765 & 10.957 &  0.014 &  0.054 &  1.255 & 23.722 &   -0.364 &   14.200 &    26821 &    5430 \\
		MSCIEM & -9.995 & 10.073 & 0.011 &  0.073 &  1.190 & 20.068 &   -0.549 &   11.163 &    14791 &    4959 \\ \hline\hline
	\end{tabular}}\caption{{\protect\footnotesize Summary statistics. `JB stat'
is the test statistic for the Jarque-Bera test of normality, with a critical
value of 5.99. `LB stat' is the test statistic for the Ljung-Box test of
serial correlation in the squared returns; the critical value based on a lag
length of 10 is 18.31. `Skewness' is the Pearson measure of sample skewness,
and `Kurtosis' a sample measure of excess kurtosis. The labels `Min' and `Max'
refer to the smallest and largest value, respectively, while `{Range}' is the
difference between these two. The remaining statistics have the obvious
interpretations. }}%
\label{tab:summaries}%
\end{table}

Treatment of the single predictive models proceeds following the steps
outlined in Section \ref{single_design}, whilst the steps outlined in Section
\ref{linear_design} are adopted for the linear predictive pool. {However, due
to the computational burden associated with the more complex models employed
in this empirical setting, we update the model parameter estimates every 50
observations only. The predictive distributions are still updated daily with
new data, with the model pool weights also updated daily using the window size
$\zeta=50$.} In the case of the S\&P500 {index}, the out-of-sample predictive
assessment is based on $\tau=3,560$ observations, while for the MSCIEM index,
the out-of-sample period comprises $\tau=3,683$ observations.

For both series, we employ three candidate predictive models of increasing
complexity: i) a na\"{\i}ve Gaussian white noise model: $\mathcal{M}_{1}:$
$y_{t}\sim i.i.d.N(\theta_{1,1},\theta_{1,2})^{\prime};$ ii) a GARCH model,
with Gaussian innovations: $\mathcal{M}_{2}:$ $y_{t}=\theta_{2,1}+\sigma
_{t}\varepsilon_{t};$ $\sigma_{t}^{2}=\theta_{2,2}+\theta_{2,3}(y_{t}%
-\theta_{2,1})^{2}+\theta_{2,4}\sigma_{t-1}^{2};$ $\varepsilon_{t}\sim
i.i.d.N(0,1);$ and iii) a stochastic volatility with jumps (SVJ) model, with
Gaussian innovations: $\mathcal{M}_{3}:$ $y_{t}=\theta_{3,1}+\exp\left(
h_{t}/2\right)  \varepsilon_{t}+\Delta N_{t}Z_{t}^{p};$ $h_{t}=\theta
_{3,2}+\theta_{3,3}h_{t-1}+\theta_{3,4}\eta_{t};$ $\left(  \varepsilon
_{t},\eta_{t}\right)  ^{\prime}\sim i.i.d.N(0,I_{2\times2});$ $Pr(\Delta
N_{t}=1)=\theta_{3,5};$ $Z_{t}^{p}\sim i.i.d.N(\theta_{3,6},\theta_{3,7}).$
The first model is obviously inappropriate for financial returns, but is
included to capture misspecification and, potentially, incompatibility. Both
$\mathcal{M}_{2}$ and $\mathcal{M}_{3}$ account for the stylized feature of
time-varying and autocorrelated return volatility, but $\mathcal{M}_{3}$ also
captures the random price jumps that are observed in practice, and is {the
only model of the three that can account for} skewness {in the predictive
distribution}. The linear predictive pool is\textbf{\ }constructed from {all
three models,} $\mathcal{M}_{1}$, $\mathcal{M}_{2}$ and $\mathcal{M}_{3}$.

For this empirical exercise we consider seven scoring rules: the log-score in
(\ref{ls_prelim}), four versions of CLS in (\ref{csr_prelim}), for the 10\%,
20\%, 80\% and 90\% percentiles, and two quantile scores (QS) evaluated at the
5th and 10th percentiles (denoted by QS 5\% and QS 10\% respectively). The QS
defined at the $p$th percentile is defined as $QS$ $p\%=(y_{t}-q_{t})\left(
\mathbf{1}_{(y_{t}\leq q)}-p\right)  ,$ with $q_{t}$ denoting the predictive
quantile satisfying $Pr(y_{t}\leq q_{t}|y_{1:t-1})=p$.\footnote{{See Gneiting
and Raftery (2007) for a discussion of the properties of QS as a proper
scoring rule.}} Use of QS (in addition to CLS) enables some conclusions to be
drawn regarding the relevance of targeting tail accuracy \textit{per se} in
the production of optimal predictions, as opposed to the importance of the
score itself. Tables \ref{tab:spx2} and \ref{tab:msciem2} report the results
for the S\&P500 and MSCIEM index respectively, with the format of both tables
mimicking that used in the simulation exercises. In particular, we continue to
use bold font to indicate the largest average score according to a given
out-of-sample measure, but now supplement this with the use of italics to
indicate the \textit{second largest} value in any column.

\begin{table}[ptbh]
\caption{{\protect\small Predictive results for the S\&P500 index returns.
Average out-of-sample scores {are recorded for the three} competing models, as
well as {for the linear} pool of these three models, based on $\tau=3,560$
out-of-sample values, covering the period between {March 16th, 2006 and May
7th, 2020}. The rows in each panel refer to the optimizer used. The columns
refer to the out-of-sample measure used to compute the average scores. The
figures in bold are the largest average scores {in each column}, while the
figures in italics are the second largest average scores.}}%
\label{tab:spx2}
\centering
\scalebox{0.9}{
		\begin{tabular}{rlccccccc}
			&  &  &  &  &  &  &  &  \\ \hline \hline
			&  &  &  &  &  &  &  &  \\
			&  & \multicolumn{7}{c}{\textbf{Out-of-sample score} }\\  \cline{3-9}
			&  &  &  &  &  &  &  &  \\
			&  & \multicolumn{1}{l}{LS} & \multicolumn{1}{l}{QS 5\%} &
			\multicolumn{1}{l}{QS 10\%} & \multicolumn{1}{l}{CLS 10\%} & \multicolumn{1}{l}{CLS 20\%
			} & \multicolumn{1}{l}{CLS 80\%} & \multicolumn{1}{l}{CLS 90\%} \\
			& \textbf{Optimizer}  &  &  &  &  &  &  &  \\ \cline{2-2}
			&  &  &  &  &  &  &  &  \\
			& LS & \textbf{-1.688} & -0.166 & \textit{-0.250} & \textbf{-0.475} & -0.726
			& -0.699 & -0.444 \\
			& QS 5\% & -2.230 & \textit{-0.165} & -0.252 & -0.594 & \textit{-0.594} & -0.966
			& -0.647 \\
			& QS 10\% & \textit{-1.769} & \textbf{-0.164} &
			\textbf{-0.244} & \textit{-0.533} & \textbf{-0.533} & -0.733 & -0.480 \\
			\multicolumn{1}{l}{\textbf{$\mathcal{M}_1$:Na\"{i}ve}} & CLS 10\% & -2.083 & -0.167 & -0.243 & -0.403 & -0.636 & -1.118 & -0.841 \\
			& CLS 20\% & -1.875 & -0.167 & -0.247 & -0.416 & -0.647 & -0.917 & -0.645 \\
			& CLS 80\% & -1.853 & -0.206 & -0.335 & -0.606 & -0.896 & \textbf{-0.653} &
			\textit{-0.397} \\
			& CLS 90\% & -2.132 & -0.271 & -0.455 & -0.853 & -1.157 & \textit{-0.654} &
			\textbf{-0.381} \\ \hline \hline
			&  &  &  &  &  &  &  &  \\
			& LS & \textit{-1.566} & -0.156 & -0.238 & -0.359 & -0.590 & -0.547 &
			\textit{-0.300} \\
			& QS 5\% & -1.884 & \textbf{-0.129} & \textit{-0.215} & -0.402 & -0.637 & -1.058
			& -0.797 \\
			& QS 10\% & \textbf{-1.468} & \textit{-0.130} &
			\textbf{-0.207} & -0.450 & -0.686 & -0.830 & -0.576 \\
			\multicolumn{1}{l}{\textbf{$\mathcal{M}_2$:GARCH}}  & CLS 10\% & -2.257 & -0.288 & -0.445 & \textbf{-0.337}
			& \textbf{-0.563} & -0.677 & -0.399 \\
			& CLS 20\% & -2.242 & -0.286 & -0.442 & \textit{-0.338} & \textit{-0.565} &
			-0.648 & -0.375 \\
			& CLS 80\% & -1.749 & -0.167 & -0.237 & -0.377 & -0.606 & \textbf{-0.543} &
			\textbf{-0.298} \\
			& CLS 90\% & -6.437 & -0.243 & -0.296 & -0.425 & -0.665 & \textit{-0.544} &
			\textbf{-0.298} \\ \hline \hline
			&  &  &  &  &  &  &  &  \\
			& LS & \textbf{-1.591} & -0.194 & \textit{-0.269} & \textit{-0.414} &
			\textit{-0.657} & \textbf{-0.669} & \textbf{-0.412} \\
			& QS 5\% & -3.110 & \textbf{-0.170} & -0.275 & -0.664 & -1.263 & -1.011 & -0.608
			\\
			& QS 10\% & -2.756 & \textit{-0.195} & \textbf{				-0.265} & -0.541 & -1.072 & -0.987 & -0.653 \\
			\multicolumn{1}{l}{\textbf{$\mathcal{M}_3$:SVJ}} & CLS 10\% & -2.351 & -0.198 & \textit{-0.269} & \textbf{				-0.413} & -0.685 & -0.990 & -0.646 \\
			& CLS 20\% & -1.870 & -0.202 & -0.275 & \textbf{-0.413} & \textbf{-0.649} &
			-0.770 & -0.497 \\
			& CLS 80\% & \textit{-1.839} & -0.218 & -0.325 & -0.493 & -0.755 & \textit{-0.676			} & -0.418 \\
			& CLS 90\% & -2.250 & -0.248 & -0.378 & -0.620 & -0.901 & -0.709 & \textit{-0.415			} \\ \hline \hline
			&  &  &  &  &  &  &  &  \\
			& LS & \textbf{-1.499} & -0.159 & -0.239 & -0.366 & -0.602 & -0.583 & -0.333
			\\
			& QS 5\% & -1.907 & \textit{-0.134} & \textit{-0.217} & -0.466 & -0.717 & -0.563
			& -0.723 \\
			\multicolumn{1}{l}{\textbf{Pooled}} & QS 10\% & \textit{-1.566} & \textbf{-0.133} &
			\textbf{-0.211} & -0.443 & -0.720 & -0.568 & -0.554 \\
			\multicolumn{1}{l}{\textbf{Forecasts}} & CLS 10\% & -2.218 & -0.266 & -0.403 & \textbf{				-0.348} & \textit{-0.577} & -0.558 & -0.473 \\
			\multicolumn{1}{l}{} & CLS 20\% & -2.111 & -0.264 & -0.401 & \textit{-0.349} &
			\textbf{-0.576} & -0.559 & -0.409 \\
			& CLS 80\% & -1.570 & -0.169 & -0.244 & -0.385 & -0.617 & \textit{-0.556} &
			\textit{-0.306} \\
			& CLS 90\% & -2.801 & -0.228 & -0.297 & -0.434 & -0.676 & \textbf{-0.555} &
			\textbf{-0.303} \\ \hline \hline
	\end{tabular}}\end{table}

\begin{table}[ptbh]
\caption{{\protect\small Predictive results for the {MSCIEM} index returns.
Average out-of-sample scores {are recorded for the three} competing models, as
well as {for the linear} pool of these three models, based on $\tau=3,683$
out-of-sample values, covering the period between {March 16th, 2006 and May
7th, 2020}. The rows in each panel refer to the optimizer used. The columns
refer to the out-of-sample measure used to compute the average scores. The
figures in bold are the largest average scores {in each column}, while the
figures in italics are the second largest average scores.}}%
\label{tab:msciem2}
\centering
\scalebox{0.9}{
\begin{tabular}{rlccccccc}
&  &  &  &  &  &  &  &  \\ \hline \hline
&  &  &  &  &  &  &  &  \\
&  & \multicolumn{7}{c}{\textbf{Out-of-sample score} }\\  \cline{3-9}
&  &  &  &  &  &  &  &  \\
&  & \multicolumn{1}{l}{LS} & \multicolumn{1}{l}{QS 5\%} &
\multicolumn{1}{l}{QS 10\%} & \multicolumn{1}{l}{CLS 10\%} & \multicolumn{1}{l}{CLS 20\%
} & \multicolumn{1}{l}{CLS 80\%} & \multicolumn{1}{l}{CLS 90\%} \\
& \textbf{Optimizer}  &  &  &  &  &  &  &  \\ \cline{2-2}
&  &  &  &  &  &  &  &  \\
& LS & \textbf{-1.692} & -0.161 & -0.243 & -0.518 & -0.789 & -0.676 & -0.450
\\
& QS 5\% & -2.332 & \textbf{-0.159} & -0.243 & -0.597 & -0.973 & -1.072 & -0.746
\\
& QS 10\% & \textit{-1.759} & \textit{-0.160} &
\textit{-0.239} & -0.557 & -0.835 & -0.707 & -0.483 \\
\multicolumn{1}{l}{\textbf{$\mathcal{M}_1$:Na\"{i}ve}} & CLS 10\% & -2.141 & -0.161 & \textbf{-0.238} & \textbf{-0.450} & \textit{-0.733} & -1.149 & -0.905 \\
& CLS 20\% & -1.890 & -0.162 & -0.241 & \textit{-0.459} & \textbf{-0.728} &
-0.906 & -0.667 \\
& CLS 80\% & -1.784 & -0.182 & -0.284 & -0.587 & -0.878 & \textbf{-0.664} &
\textit{-0.439} \\
& CLS 90\% & -1.985 & -0.227 & -0.368 & -0.762 & -1.068 & \textit{-0.668} &
\textbf{-0.431} \\ \hline \hline
&  &  &  &  &  &  &  &  \\
& LS & \textbf{-1.604} & -0.152 & -0.234 & -0.400 & -0.664 & -0.580 & -0.358
\\
& QS 5\% & -1.922 & \textbf{-0.125} & \textit{-0.213} & -0.405 & -0.681 & -1.066
& -0.815 \\
& QS 10\%  & -1.986 & \textit{-0.129} & \textbf{-0.208} & -0.423 & -0.703 & -1.093 & -0.852 \\
\multicolumn{1}{l}{\textbf{$\mathcal{M}_2$:GARCH}} & CLS 10\% & -2.439 & -0.326 & -0.499 & \textbf{-0.383}
& \textit{-0.650} & -0.810 & -0.557 \\
& CLS 20\% & -2.340 & -0.302 & -0.466 & \textbf{-0.383} & \textbf{-0.647} &
-0.727 & -0.483 \\
& CLS 80\% & \textit{-1.704} & -0.160 & -0.237 & -0.414 & -0.680 & \textbf{-0.575} & \textbf{-0.355} \\
& CLS 90\% & -1.715 & -0.160 & -0.239 & -0.413 & -0.682 & \textbf{-0.575} &
\textbf{-0.355} \\ \hline \hline
&  &  &  &  &  &  &  &  \\
& LS & \textbf{-1.689} & \textit{-0.202} & \textit{-0.278} & \textbf{-0.462}
& \textbf{-0.734} & \textit{-0.699} & -0.471 \\
& QS 5\% & -2.630 & \textbf{-0.181} & -0.307 & -0.703 & -1.200 & -1.004 & -0.707
\\
& QS 10\% & -2.409 & -0.203 & \textbf{-0.268} &
-0.560 & -0.972 & -0.966 & -0.684 \\
\multicolumn{1}{l}{\textbf{$\mathcal{M}_3$:SVJ}} & CLS 10\% & -2.356 & -0.207 & -0.281 & \textit{-0.471}
& \textit{-0.764} & -0.951 & -0.702 \\
& CLS 20\% & \textit{-1.848} & -0.211 & -0.288 & -0.483 & -0.765 & -0.764 &
-0.528 \\
& CLS 80\% & -2.038 & -0.215 & -0.306 & -0.537 & -0.818 & \textbf{-0.697} &
\textit{-0.466} \\
& CLS 90\% & -2.494 & -0.260 & -0.374 & -0.669 & -0.980 & -0.717 & \textbf{-0.462} \\ \hline \hline
&  &  &  &  &  &  &  &  \\
& LS & \textbf{-1.539} & -0.149 & -0.232 & -0.410 & -0.626 & -0.594 & -0.372
\\
& QS 5\% & -1.921 & \textbf{-0.127} & \textit{-0.214} & \textit{-0.409} & -0.631
& -0.595 & -0.752 \\
\multicolumn{1}{l}{\textbf{Pooled}} & QS 10\% & -1.944 & \textit{-0.133} & \textbf{-0.211} & -0.415 & -0.631 & -0.599 & -0.751 \\
\multicolumn{1}{l}{\textbf{Forecasts}} & CLS 10\% & -2.320 & -0.277 & -0.416 & \textbf{-0.392} & \textit{-0.582} & -0.588 & -0.630 \\
& CLS 20\% & -1.853 & -0.154 & -0.236 & -0.421 & \textbf{-0.429} & -0.626 &
-0.656 \\
& CLS 80\% & \textit{-1.571} & -0.158 & -0.239 & -0.430 & -0.665 & \textbf{-0.586} & \textit{-0.361} \\
& CLS 90\% & -1.601 & -0.160 & -0.246 & -0.443 & -0.681 & \textbf{-0.586} &
\textbf{-0.360} \\ \hline \hline
\end{tabular}}\end{table}

We make three comments regarding the empirical results. {\textit{First}, for
both data set}s, and for all three single models, strict coherence is close to
holding uniformly, with most of the diagonal elements in all panels being
either the highest (bolded) or the second highest (italics) values in their
respective columns. This suggests that each individual model, whilst
inevitably a misspecified version of the true unknown DGP, is compatible
enough with the true process to enable score-specific optimization to reap
benefits. \textit{Second}{, we remark th}at all three individual models are
quite distinct, and are likely to be associated with quite different degrees
of estimation error. Hence, while the na\"{\i}ve model is no doubt the most
misspecified, given the documented features of both return series, it is also
the most parsimonious and, hence, likely to produce estimated scores with
small sampling variation. Thus, it is difficult to assess wh{ich model has the
best predictive performance overall, due to the interplay between sampling
variation and model misspecification (see \citealp{patton2019comparing}, for
an extensive investigation of this issue). While the matter of model selection
\textit{per se }is not the focus of the paper, we do note that of the single
models, the Gaussian GARCH(1,1) model estimated using the }%
relevant{\textbf{\ }score-specific optimizer }is the best performer
out-of-sample overall, according to all measures.\ \textit{Third, }{we note
that the pooled foreca}sts exhibit close to uniform strict coherence, for both
series, highlighting that the degree of misspecification in the pool is still
sufficient for benefits to be had via score-specific optimization. However,
the numerical gains reaped by score-specific optimization in the case of the
pool\textbf{ }are \textit{typically} not as large\textbf{ }as in the single
model cases. That is, and as is consistent with earlier discussion, the
additional flexibility produced by the pooling\textbf{ }can reduce the
ability\textbf{ }of score-specific optimization to produce marked predictive
improvements in some instances.\footnote{For both data sets the (time-varying)
weights in the linear pool (not recorded here for reasons of space) tend to
favour the GARCH(1,1) model most frequently. This finding is consistent with
the fact that the magnitudes of the average scores for the linear pool are
most similar to the corresponding values for the GARCH(1,1) model.}

\section{Discussion\label{disc}}

{This paper contributes to a growing literature in which the role of scoring
rules in the production of \textit{bespoke} forecasts - i.e. forecasts
designed to be optimal according to a particular measure of forecast accuracy
- is given attention. With our focus solely on probabilistic forecasts, our
results highlight the care that needs to be taken in the production and
interpretation of such forecasts. It is not \textit{assured} that optimization
according to a problem-specific scoring rule }will {yield benefits; the
relative performance of so-called `optimal' forecasts depending on the nature
of, and interplay between: the true model, the assumed model and the score.
}That is, if the predictive model simply does not allow a given score to
reward the type of accuracy it should, optimization with respect to that score
criterion comes to naught.{\textbf{\ }}One may as well use the simplest
optimizer for the problem at hand, and leave it at that.\ However, subject to
a{\ basic match, or compatibility, between the true process and the assumed
predictive model, it is certainly the case that optimization \textit{can
}produce accuracy gains in the manner intended, with the gains being more
marked the greater the degree of misspecification.}

{Knowing \textit{when} optimization will yield benefits in any particular
empirical scenario is difficult, but the use of a plausible predictive model
that captures the key features of the true data generating process is
obviously key. The results in the paper also highlight the fact that use of
score-specific optimization in the linear pool context is likely to reap less
benefits than in the context of a single misspecified model. Theoretical
exploration and characterization of all of these matters is likely to prove
difficult, given the number of aspects at play; however such work, even if
confined to very \textit{specific} combinations of generating
process/model/scoring rule, would be of value. We leave such explorations for
future work.}

\bibliographystyle{apalike}
\bibliography{Biblio}

\end{document}